\documentclass[11pt, letterpaper]{article}

\usepackage[english]{babel} 
\usepackage{amssymb}
\usepackage{amsmath}

\usepackage{txfonts}
\usepackage[classicReIm]{kpfonts}

\usepackage{mathdots}
\usepackage{graphicx} 

\usepackage{latexsym}
\usepackage{hanging}
\usepackage{setspace}
\usepackage{geometry}
\usepackage[compact]{titlesec} 
\titlespacing{\section}{0pt}{*0}{*0}
\usepackage{array}
\newcolumntype{L}{>{\centering\arraybackslash}m{3.3cm}}
\newcolumntype{M}{>{\centering\arraybackslash}m{3.6cm}}
\newcolumntype{N}{>{\centering\arraybackslash}m{3.2cm}}
\newcolumntype{T}{>{\centering\arraybackslash}m{3.8cm}}
\newcolumntype{P}{>{\centering\arraybackslash}m{6.0cm}}
\newcolumntype{Q}{>{\centering\arraybackslash}m{4.2cm}}
\newcolumntype{A}{>{\arraybackslash}m{1.9cm}}
\newcolumntype{B}{>{\arraybackslash}m{0.9cm}}
\usepackage{multirow}
\usepackage{array, boldline, makecell, booktabs}

\usepackage[colorlinks=true]{hyperref}
\usepackage{authblk}

\title{Form + Function: Optimizing Aesthetic Product Design via Adaptive, Geometrized Preference Elicitation}

\author{Namwoo Kang \thanks{nwkang@sookmyung.ac.kr, Mechanical Systems Engineering, Sookmyung Women's University}, Yi Ren \thanks{yiren@asu.edu, Mechanical Engineering, Arizona State University. First two authors have equal contributions.}, Fred Feinberg \thanks{feinf@umich.edu, Ross School of Business, University of Michigan, Ann Arbor}, and Panos Papalambros \thanks{pyp@umich.edu, Mechanical Engineering, University of Michigan, Ann Arbor}}
\date{}                     
\begin{document}
\maketitle

\begin{center}
\textbf{Abstract}
\end{center}

Visual design is critical to product success, and the subject of intensive 
marketing research effort. Yet visual elements, due to their holistic and 
interactive nature, do not lend themselves well to optimization using extant 
decompositional methods for preference elicitation. Here we present a 
systematic methodology to incorporate interactive, 3D-rendered product 
configurations into a conjoint-like framework. The method relies on rapid, 
scalable machine learning algorithms to adaptively update product designs 
along with standard information-oriented product attributes. At its heart is 
a parametric account of a product's geometry, along with a novel, adaptive 
``bi-level'' query task that can estimate individuals' visual design form 
preferences and their trade-offs against such traditional elements as price 
and product features. We illustrate the method's performance through 
extensive simulations and robustness checks, a formal proof of the bi-level 
query methodology's domain of superiority, and a field test for the design 
of a mid-priced sedan, using real-time 3D rendering for an online panel. 
Results indicate not only substantially enhanced predictive accuracy, but 
two quantities beyond the reach of standard conjoint methods: trade-offs 
between form and function overall, and willingness-to-pay for specific 
design elements. Moreover -- and most critically for applications -- the 
method provides ``optimal'' visual designs for both individuals and 
model-derived or analyst-supplied consumer groupings, as well as their 
sensitivities to form and functional elements. \\

\noindent{\textbf{Keywords}: product design optimization; conjoint analysis; machine 
learning; preference elicitation; visual design}

\newpage 

\section{Introduction}
\label{subsec:mylabel1}

A product's visual form has long been acknowledged as a pivotal element of 
consumer choice (Kotler and Rath 1984, Bloch 1995, Veryzer and Hutchinson 
1998, Dahl et al. 1999, Bloch et al. 2003, Pan et al. 2017). Firms as 
diverse as Bang {\&} Olufsen, Apple, Dyson, Uniqlo, and Tesla have not only 
made striking visuals an emblematic linchpin of their success, but have also 
helped propel design, as both discipline and corporate mission, into the 
public sphere. Even old-line firms have taken note: IBM's product design 
initiative is now the world's largest, and HP relishes its rivalry with 
Apple in that sphere.\footnote{ IBM Design (www.ibm.com/design/teams/); Paste Magazine (www.pastemagazine.com/articles/2017/08/the-story-of-spectre-how-hp-reinvented-itself-thro.html)}

Academic research in both marketing and engineering has consequently 
grappled with how to capture, and ultimately optimize, product ``form,'' 
with varying definitions, terminologies, and degrees of success. Preference 
modeling studies incorporating visual design elements have addressed general 
concepts like ``appearance'' (Creusen and Schoormans 2005), ``form'' (Bloch 
1995, Orsborn and Cagan 2009, Tseng et al. 2012), ``styling'' (Dotson et al. 
2019), and ``design'' (Landwehr et al. 2011, Burnap et al. 2019), as well as 
more specific ones like ``shape'' (Orsborn et al. 2009, Kelly et al. 2011, 
Orbay et al. 2015), ``silhouette'' (Reid et al. 2012), and ``profile'' (Lai 
et al. 2005). ``Form'' -- broadly construed -- often assumes a central role 
in real-world preference modeling problems. According to Bloch (1995), form 
helps attract consumer attention, communicate product information, and 
stimulate visual pleasure, thereby generating a long-lasting perceptual 
impression. In practice, sales predictions can be significantly improved by 
accommodating form (Landwehr et al. 2011). Marketers and designers also find 
valuable trade-offs between form and functional attributes (Dotson et al. 
2019, Sylcott et al. 2013a), and revealing such trade-offs can lead to 
superior balance between visual appeal and functionality (Reid et al. 2012).

Product aesthetics comprises purely visual elements like color and 
packaging, haptic (Peck and Childers 2003) and sensory (Krishna 2012) 
aspects that can be altered and optimized independently, and more 
nuts-and-bolts `geometric' elements that both convey product image and 
constrain / interact with the internal operations of the product itself. 
Honing these geometrical elements is critical for efficient design of 
components and production processes, but mapping from the geometry of a 
product to how much potential consumers might like it is a complex exercise 
(Michalek et al. 2005, 2011). A further challenge is quantifying how 
elements of form preference are traded off against `traditional' attributes 
like price and performance. Such trade-offs are critical in allowing 
designers to determine not only how important ``design'' is overall to a 
particular consumer or type of consumer, but whether specific \textit{aspects} of design -- 
like a low-profile car hood that may constrain the powertrain and require 
costly amendments to maintain crashworthiness -- can be accommodated, 
subject to supply-side and consumer budget constraints.

Marketers have traditionally employed conjoint-based techniques to assess 
consumer trade-offs (Green and Srinivasan 1990; Green, Krieger, and Wind 
2001), but have struggled to incorporate product form into overall 
preference modeling. One method of doing so involves the elicitation of 
adjectival descriptors. For example, the largest survey of its kind, Maritz 
Corp.'s New Vehicle Car Survey (NVCS), asks 250,000 new car buyers annually 
to rate statements like ``styling is at or near the top of important 
characteristics in a new vehicle''; to rate the importance of, satisfaction 
with, and likelihood of rejecting vehicles based on interior and exterior 
styling; to assess trade-offs with vehicle safety; and to assess 28 
``image'' elements.\footnote{ Specifically, the 28 image elements were: 
classic, responsive, youthful, bold, luxurious, prestigious, stylish, 
functional, safe, innovative, economical, simple, conservative, 
environmentally-friendly, sleek, distinctive, fun to drive, comfortable, 
well-engineered, rugged/tough, elegant, family-oriented, good value, sporty, 
powerful, sophisticated, exciting.} Yet this procedure still leaves 
designers in a quandary as to \textit{what to actually design}: while it's helpful to learn a particular 
consumer likes ``sleek'' cars, what does such a car actually look like? 
Reasonable people can differ on the operationalization of adjectival labels: 
one consumer's sleek is another's clunky, and a sleek SUV may have a 
radically different visual footprint from a sleek sports car. Moreover, 
potential buyers (or segments thereof) can express design preferences but be 
in no position to enact them, due to financial, familial, or other 
constraints, making it difficult to know how ``design,'' as an overall 
attribute -- or various aspects of design, like ``ruggedness'' -- is 
traded-off against more prosaic but measurably important elements like 
price, safety, and other features.

We later review research in preference modeling, efficient algorithms, and 
design optimization; yet, to the authors' knowledge, few studies 
combine a \textit{form} preference model with one for \textit{overall} preference -- that is, including 
``traditional'', conveyable product attributes -- and even these few rely on 
decomposition via disjoint measurement (Sylcott et al. 2013a, Dotson et al. 
2019). This is reasonable for lab studies where strict demographics controls 
can be exacted; it is more problematic when applied to online 
``crowdsourced'' groups that can differ in crucial characteristics. 
Combining or ``fusing'' data across such groups presents serious impediments 
to accurately capturing individual-level preference (Feit et al. 2010, Feit 
et al. 2013, Qian and Xie 2013).

To overcome these challenges, we propose a new methodology that disentangles 
\textit{form preference} from \textit{overall preference} and coordinates them adaptively, in real time, while allowing 
respondents to manipulate 3D renderings of product geometry. The method 
makes use of both Bayesian methods for preference measurement and machine 
learning tools for rapidly adapting a multidimensional space whose 
\textit{consumer-oriented visual design }characteristics (e.g., boxy, retro, etc.) do not need to be determined in 
advance, or indeed at any point via verbal protocols. We apply the proposed 
``bi-level adaptive conjoint'' method to model both form preference (for 
vehicles) based on underlying product geometry and overall preference by 
revealing trade-offs between form (e.g.,~hood length, windshield pitch) and 
functional attributes (here, price and fuel efficiency). Results -- via both 
Monte Carlo simulation and a crowdsourced experiment -- suggest that the 
proposed method not only zeroes in on each consumer's preferred visual 
design, but, more generally, can both elicit superior individual-level 
preference estimates than conventional conjoint alone and produce 
``optimal'' designs for analyst-supplied consumer groupings. 

These benefits in turn allow, we believe for the first time, explicit 
measurement of trade-offs among willingness-to-pay (WTP) and design 
variables, e.g., whether price-sensitive consumers have marked preferences 
for certain styles, whether sports car buyers are less concerned about fuel 
economy, etc. Although the bi-level adaptive query method is novel, the 
set-up is `modular' in the sense of the ability to slot in other question 
types, although their efficiency, scalability, and speed would need careful 
testing. Rather, as we emphasize throughout, the key innovation is allowing 
consumers to interact with a parametric geometrization of the product's 
topology, within a conjoint-like framework, to enable real-time, individual- 
or group-level design optimization.

The paper is structured as follows: Section 2 reviews related approaches 
from both the marketing and engineering design literatures. Section 3 
proposes the ``bi-level adaptive conjoint'' method, whose advantages over 
traditional conjoint are illustrated both by extensive simulations in 
Section 4 and web-based panelists in Section 5. Section 6 concludes by 
discussing findings and potential extensions.

\section{Prior Literature on Product Design Preference Optimization}
\label{subsec:mylabel2}
The literature on product design, both as a stand-alone field and within 
cognate disciplines, is vast. We therefore provide concise integrative 
discussions of three lines of prior research directly bearing on the 
subsequent development: Section 2.1 addresses methods for eliciting form 
preference only; Section 2.2 discusses eliciting both form preference and 
overall preference; and Section 2.3 reviews optimization and machine 
learning algorithms for preference elicitation generally and conjoint 
analysis specifically.

\subsection{Form Preference Modeling}
\label{subsec:mylabel3}
Form preference has been addressed in engineering design using a variety of 
approaches, mainly differing in terms of the parametric nature of the 
preference function, and secondarily in terms of how these parameters are 
estimated (although we later discuss estimation extensively for the 
real-time adaptive portion of our implementation, we will be largely 
agnostic regarding estimation technologies otherwise). Table 1 summarizes 
previous research focused on eliciting form preference via parametric 
models, primarily from the engineering design field.
\begin{table}[]
    \centering
    \small{
    \begin{tabular}{p{1.7cm} p{2.2cm} p{1.9cm} p{1.5cm} p{1.4cm} p{1.4cm} p{1.2cm} p{1.6cm}}
        \hline
        \textbf{Research} & \textbf{Parametric preference function}	& \textbf{Parameter estimation}	& \textbf{Hetero\-geneity}	& \textbf{Survey} & \textbf{Query design}	& \textbf{Product} & \textbf{Repre\-sentation} \\
        \hline
        Lai et al. (2005)	& S/N ratio	& Taguchi	& Aggregate	& Rating & Non\-adaptive &	Vehicle	& 2D \quad\quad Silhouette \\ \hline
        Orsborn et al. (2009) &	Quadratic &	BTL &	Individual & Choice	& Non\-adaptive &	Vehicle	& 2D \quad\quad Silhouette \\ \hline
        Kelly et al. (2011)	& Quadratic w/
        interaction & PREFMAP &	Aggregate &	Rating &	Non\-adaptive	& Water bottle & 2D \quad\quad Silhouette \\ \hline
        Lugo et al. (2012) & Linear	& Regression &	Aggregate &	Rating & Non\-adaptive &	Wheel rim & 2D \quad\quad Rendering \\ \hline
        Reid et al. (2012) & Linear	& Regression &	Aggregate & Rating & Non\-adaptive &	Vehicle	& 2D \quad\quad Silhouette \\ \hline
        Tseng et al. (2012)	& Neural \quad\quad network &	ANN	& Aggregate	& Rating & Non\-adaptive &	Vehicle	& 2D \quad\quad Silhouette \\ \hline
        Reid et al. (2013) & Linear	& BTL &	Aggregate & Choice & Non\-adaptive &	Vehicle \& carafe	& 3D \quad\quad Rendering \\ \hline
        Sylcott et al. (2013) &	Linear with interaction term & MNL & Aggregate & Choice & Non\-adaptive & Vase \& vehicle & 2D \quad\quad Silhouette \\ \hline
        Sylcott et al. (2016) & Linear & MNL &	Aggregate & Choice & Non-adaptive & Knife & 3D Printed \\ \hline
        Pan et. al (2017) &	Neural \quad\quad network &	Adversarial Training & Individual & Choice & Non\-adaptive & Vehicle	& Pixels (50K) \\ \hline
        Burnap et. al (2019) & Neural \quad\quad network & VB + \quad\quad Adversarial & Aggregate & Rating & Non\-adaptive & Vehicle & Pixels (200K) \\ \hline
        Dotson et al. (2019) & Linear & MNP & Individual & Choice + Rating & Non\-adaptive & Vehicle & 2D \\
        \hline
    \end{tabular}}
    \caption{Parametric Models for Eliciting Form Preference}
    \label{tab:tab1}
\end{table}
Among the first approaches to optimizing the visual design space was Lai, 
Chang, and Chang (2005), who tested 2D designs for a passenger car by having 
three professional product designers develop appropriate initial candidates, 
which were then broadened using images of 125 existing passenger cars, and 
subsequently culled by a panel of experts to 27 combinative designs for a 
(parametric) Taguchi experiment. Despite its pioneering nature in 
calibrating form preference, their study would be of limited interest to 
marketers, due to its lack of parametric heterogeneity, the non-adaptive 
nature of its querying strategy, and to a lesser extent its reliance on 2D 
models and a ratings-based conjoint approach. Lugo et al. (2012) designed a 
wheel rim, and Reid et al. (2013) a vehicle shape, using similar methods, 
with a linear preference function estimated via standard regression 
techniques, but also adaptively at the aggregate level for 2D designs. By 
contrast, quadratic preferences that allowed for internal extrema were 
applied by Orsborn et al. (2009) to 2D shape design, using a choice-based 
instrument at the individual level. Sylcott et al. (2013b) included 
interaction terms among attributes, and Kelly et al. (2011) allowed both 
quadratic preferences and potential interactions, although both were limited 
to aggregated inferences. Interaction terms are especially important for 
form optimization, since some consumer-valued qualities depend on multiple 
geometrical elements: for example, a product is only ``compact'' through an 
interaction among its dimensions.

There are several limitations in this line of engineering design research. 
First, as noted earlier, most has sidestepped preference heterogeneity, so 
that results would suggest a single ``optimal'' design suitable for the 
population as a whole. Second, with few exceptions this line of work has 
relied on non-adaptive (and sometimes non-choice-based) query designs, which 
are demonstrably outperformed by adaptive techniques (Toubia et al. 2003, 
Toubia et al. 2004, Abernethy et al. 2008), which we take up again in 
Section 2.3 and in our proposed method. Third, nearly all prior research in 
the area has relied on 2D product representations, although there are 
exceptions that did not explicitly calibrate a form preference model (e.g., 
Ren and Papalambros 2011, Reid et al. 2013). In the marketing literature, 
Kim, Park, Bradlow, and Ding (2014) emphasized the importance of different 
kinds of attributes and ran conjoint studies that involved product designs 
for both hedonic (e.g., sunglasses) and utilitarian (e.g., coffeemakers) 
products, using 2D designs from a candidate list of 20 possibilities, in a 
manner similar to Dotson et al. (2019). To control 3D rendering 
parametrically is a challenge for optimization algorithms (e.g.,~Hsiao and 
Liu 2002) and conjoint interface designers, as well as for participants who 
may find such representations cumbersome to navigate, despite their being 
perceptually essential.

More recently, machine learning architectures -- for example, generative 
adversarial networks -- have been employed to process of large corpora of 2D 
imagery, particularly for vehicle design (Pan et al. 2017, Burnap et al. 
2019). This approach allows ``learning'' of essentially arbitrary degrees of 
complexity and interaction among the (2D) design elements, but depends on a 
supervised (i.e., labeled) set of stimuli to interrelate design elements and 
aesthetic preference(s). Such methods can scale to large stimuli sets; for 
example, Pan et al. (2017) modeled perceptions of four pairs of aesthetic 
design attributes (e.g., ``Sporty'' vs. ``Conservative'') for over 13,000 2D 
images of SUVs from 29 manufacturers, while Burnap et al. (2019) analyzed 
aesthetic image ratings from 178 consumers who participated in ``theme 
clinics'' to evaluate 203 unique SUVs. Labeled stimuli show great promise as 
a way to engage in pre-study image analysis. By contrast, our use of machine 
learning focuses on adaptively altering each respondent's 3D design in 
real-time, and to concurrently evaluate trade-offs between form and 
function.

\subsection{Joint Modeling of Form and Overall Preference }
\label{subsec:mylabel4}
Engineers, industrial designers, and marketers often face conflicting design 
choices due to trade-offs between form and function attributes. For example, 
consumers typically want products -- cell phones, clothing, automobiles, 
etc. -- that are trim (form) yet durable (function), or sophisticated (form) 
yet moderately priced (function), attributes with conflicting design 
imperatives. There is presently little formal research to guide this 
trade-off. Two papers other than the present one have specifically addressed 
it, both centering on vehicle design; critical differences in approach and 
data requirements among the three studies are summarized in Table 2.
\begin{table}[]
    \centering
    \small{
    \begin{tabular}{p{4.5cm} L M N}
    \hline
    & \textbf{Dotson et al. (2019)}	& \textbf{Sylcott et al. (2013a)}	& \textbf{This study} \\ \hline
    \textbf{Survey}	& Two separate surveys
 \quad (1) form: rating \quad \quad \quad (2) overall: choice & Three separate surveys (1) form: choice \quad\quad \quad (2) function: choice \quad \quad \quad (3) overall: pairwise comparison & Bi-level questions in single survey \quad \quad (1) form: metric paired-comparison \quad \quad (2) overall: choice \\
 \hline
\textbf{Time delay between surveys}	& Yes & Yes & No (real time) \\ \hline
\textbf{Query design} & Non-adaptive & Non-adaptive & Adaptive \\ \hline
\textbf{Preference function} & Form: covariance structure \quad \quad \quad Overall: linear &	Form: quadratic
Function: linear \quad \quad \quad Overall: linear &	Form: radial basis
Overall: linear \\ \hline
\textbf{Estimation}	& Form: Euclidian distance
\quad \quad \quad Overall: Bayesian &	Bradley-Terry-Luce (BTL)	& Form: Rank SVM mix 
\quad \quad \quad \quad Overall: HB \\ \hline
\textbf{Heterogeneity}	& Individual & Individual	& Individual \\
\textbf{Product}	& Vehicle	& Vehicle	& Vehicle \\
\hline
    \end{tabular}}
    \caption{Approaches to Relating Form and Function Preferences}
    \label{tab:tab2}
\end{table}
One approach is to simply measure form and overall preference separately, 
then knit them together in a model-based manner. In this vein, Sylcott 
et al. (2013a) conducted three separate conjoint surveys: the first for form 
preference; the second for function preference (without price); and the last 
for overall preference using two meta-attributes (e.g., form and function, 
each with three levels like low, medium, and high). Despite its intricacy, 
this approach not only potentially exacerbates problems stemming from 
demographic differences among sampled groups, it precludes incorporating 
form into an \textit{overall} preference model, leaving their relationship indeterminate. 
Specifically, zeroing in on the individual-level sweet spot in the joint 
space of product designs and traditional conjoint attributes is 
impracticable.

Dotson et al. (2019) model the effect of 2D images on product choice by 
accounting for the utility correlation between similar-looking images and 
augmenting the standard choice-based conjoint task with direct ratings of 
image appeal. ``Form'' is accommodated in overall preference by including 
(population) mean image appeal ratings a separate attribute in the product 
utility specification; and heterogeneity in image appeal is modeled via a 
multinomial probit where the utility error covariance depends on the 
correlations in consumer image ratings. Despite the inherent scalability of 
the approach, it relies on collecting additional ratings data, and assuming 
that the underlying Euclidean distance metric for control points on the 2D 
product (automobile) shape captures \textit{consumers'} perceptions of design similarity. Our 
proposed method, by contrast, obviates the need for separate pictorial 
evaluations, that it's possible to capture ``shape distance'' using a single 
(distance) metric, or indeed presuming that 2D imagery suffices to represent 
product topology. Rather, ``form'' and ``function'' information is collected 
contemporaneously, from the same individuals, and thereby allows an 
assessment of design preference heterogeneity unrelated to preference 
scales, metrics, or pre-standardized imagery.

This hints at a deeper problem bedeviling design optimization methodology 
overall: reliance on large stimuli sets is exacerbated by the intrinsic 
nonlinearity of the visual design space. For example, a consumer who likes 
pick-ups and loves sports cars may dislike mid-sized sedans, despite their 
being ``intermediate'' in a canonical parameterization of ``car shape 
space''. That is, designers cannot determine a set of stimuli among which a 
target consumer has modest preference differentials and simply interpolate 
between them to determine a utility surface. Moreover, keeping target visual 
stimuli constant across respondents runs the risk of mismeasuring 
heterogeneity: ideally, each respondent should be able to veer into the 
region of the design space containing her most preferred product 
configuration, as opposed to being imprisoned within the convex hull or 
along the simplex edge bordered by the pre-configured designs. Lastly, 
screen images used in prior research are necessarily two-dimensional and 
fixed, projections of the 3D designs that respondents must visualize and 
integrate to fully experience: Although this reduces both complexity in 
presentation and latency in administration, it substantially reduces realism 
and respondent involvement. Our methodology therefore works entirely with 
configurable, on-the-fly renderable, 3D product representations, with no 
prior or exogenous human evaluation.

\subsection{Optimization and Machine Learning Algorithms}
\label{subsec:mylabel5}
The approach developed here leverages advances in computing power, web-based 
query design, machine learning, and optimization to allow efficient, 
scalable, real-time form optimization, as opposed to simply choosing among a 
pre-determined set of (typically 2D) alternatives. For purposes of 
comparison, we summarize research according to two dimensions: estimation 
methods / shrinkage properties (Table 3), and adaptive query design (Table 
4). Our coverage is again selective and deliberately concise. The interested 
reader is referred to Toubia, Evgeniou, and Hauser (2007; their Table 1 
especially) for extensive background on methods; to Netzer et al. (2008) for 
general challenges in preference elicitation; to Toubia, Hauser, and Garcia 
(2007) for both simulation results and empirical comparisons among 
traditional and polyhedral methods; to Halme and Kallio (2011) for an 
overview of choice-based estimation; to Chapelle et al. (2004) for machine 
learning in conjoint; to Dzyabura and Hauser (2011) specifically in 
reference to adaptive questionnaire design; to Huang and Luo (2016) for 
``fuzzy'' and SVM-based complex preference elicitation; and to Burnap et al. 
(2019) for supervised learning approaches to aesthetics. 
\begin{table}[]
    \centering
    \small{
    \begin{tabular}{T P c}
    \hline
    \textbf{Research} & \textbf{Method} & \textbf{Shrinkage} \\
    \hline
    Lenk et al. (1996)
Rossi and Allenby (2003) & Hierarchical Bayes & 	 Yes \\ \hline
Toubia et al. (2003) & Metric paired-comparison analytic-center estimation & No \\ \hline
Toubia et al. (2004) & Adaptive choice-based
analytic-center estimation 	&  No \\ \hline
Cui and Curry (2005) & Support Vector Machine (SVM) &	 No \\ \hline
Evgeniou et al. (2005)	& SVM mix &	 Yes \\ \hline
Evgeniou et al. (2007)	& Heterogeneous partworth estimation
with complexity control &	 Yes \\ \hline
Dzyabura and Hauser (2011)	& Variational Bayes Active Learning
with Belief Propagation &	 Yes \\ \hline
Burnap et al. (2016)	& Restricted Boltzmann Machine and Convex Low-Rank Matrix Estimation & 	Yes \\ \hline
Huang and Luo (2016) &	Fuzzy SVM estimation & Yes \\ \hline
This study 	& Form preference: Rank SVM mix
Overall preference: Hierarchical Bayes	& Yes \\ \hline

    \end{tabular}}
    \caption{Estimation Methods for Preference Elicitation Models}
    \label{tab:tab3}
\end{table}

Hierarchical Bayesian (HB) methods have long been popular for estimating 
partworths in conjoint, and serve to overcome sparse individual-level 
information by shrinkage towards a population-based density (Lenk et al. 
1996, Rossi and Allenby 2003). Due to stability and its near-ubiquitous use 
in applications (e.g., Sawtooth), we adopt HB for estimating 
individual-level partworths of the overall preference model. Toubia et al. 
(2003) proposed a polyhedral method especially well-suited to metric 
paired-comparison query design; this has been extended to adaptive choice 
queries using classical and Bayesian approaches (Toubia et al. 2004; Toubia 
and Flores 2007). In a similar vein, Evgeniou et al. (2007) proposed a 
distinct (compared with HB) method for shrinking individual-level shrinkage, 
minimizing a convex loss function.

Cui and Curry (2005) and Evgeniou et al. (2005) extended the Support Vector 
Machine (SVM) -- a popular machine learning algorithm used in classification 
problems -- to conjoint applications. Specifically, Evgeniou et al. (2005) 
proposed an SVM mix that can accommodate parametric heterogeneity by 
shrinking individual-level partworths toward population-based values using a 
linear sum. Because it lowers computational cost dramatically compared to 
HB, at a comparable level of accuracy, we adapt this method for form 
preference as well as adaptive design for both form and overall queries. As 
detailed in Section 3, we couple this with a Gaussian kernel rank SVM mix to 
handle non-linear form preference.
\begin{table}[]
    \centering
    \small{
    \begin{tabular}{N Q Q N} 
    \hline
        \textbf{Research} & \textbf{Method} & \textbf{Sampling} & \textbf{Data used} \\
    \hline
Toubia et al. (2003) & Adaptive metric paired-comparison polyhedral question design &	Minimize polyhedron volume and length of longest axis & Individual's prior responses \\ \hline
Toubia et al. (2004) Toubia, Hauser, and Garcia (2007)	& Adaptive choice-based polyhedral question design & Minimize polyhedron volume and length of longest axis & Individual's prior responses \\ \hline
Abernethy et al. (2008)	& Hessian-based adaptive choice-based conjoint analysis & Maximize smallest positive eigenvalue of loss function Hessian & 	Individual's prior responses \\ \hline
Dzyabura and Hauser (2011) & Active machine learning (Adaptive) + Variational Bayes	& Maximize expected information gain (reduction in posterior entropy) & Synthetic +  Individual Responses \\ \hline
Huang and Luo (2016) & Adaptive Fuzzy SVM; collaborative filtering	& Tong and Koller (2001) + minimal ratio margin	& Both individual's and others' prior responses \\ \hline
This study & Adaptive metric paired-comparison SVM mix question design & Minimize difference between utilities of new pairs and maximize Euclidean distance among all profiles	& Both individual's and others' prior responses \\ \hline
    \end{tabular}}
    \caption{Adaptive Query Design Methods }
    \label{tab:tab4}
\end{table}

Adaptive question design methods for conjoint analysis typically employ 
``utility balance''$^{\, }$(Huber and Zwerina 2007, Abernethy et al. 2008), 
wherein profiles in each choice set have similar utilities based on 
partworths estimated from previous answers (Toubia et al. 2007a, p. 247); 
the approach is comparable to ``uncertainty sampling'' for query strategy in 
the machine learning field (Settles 2010). Previous research on adaptive 
querying, outlined in Table 4, has demonstrated that it generally 
outperforms non-adaptive designs, especially when response errors are low, 
heterogeneity is high, and the number of queries is limited (Toubia et al. 
2007a). 

Polyhedral methods (e.g., Toubia et al. 2003, 2004, 2007) typically select a 
next query by minimizing polyhedral volume longest axis length, thereby 
finding the most efficient constraints (i.e., queries) to reduce the 
uncertainty of feasible estimates (i.e., the polyhedron). By contrast, 
Abernethy et al. (2008) measure uncertainty as the inverse of the Hessian of 
the loss function and select the next query by maximizing its smallest 
positive eigenvalue. More recently, Huang and Luo (2016) -- building on 
methods from Tong and Koller (2001), Lin and Wang (2002), and Dzyabura and 
Hauser (2011) -- proposed an adaptive decompositional framework for high 
dimensional preference elicitation problems, using collaborative filtering 
to obtain initial partworths, must-haves and/or unacceptable features. They 
also leverage previous respondents' data, generating questions using fuzzy 
SVM active learning and utility balance, as we do.$^{\, }$\footnote{ Hauser 
and Toubia (2005) demonstrate that imposing utility balance can lead to 
biased partworths and ratios thereof, in a metric conjoint setting. While we 
cannot claim our forthcoming bi-level query task is immune to such biases, 
it consists of alternating (4-point, symmetric) ordinal and dichotomous 
choice questions, and we later empirically examine the impact of various 
cutoff values for the former, finding them to have minimal effect on 
predictive accuracy. It has been shown theoretically that active learning 
(adaptive sampling) requires a balance between exploitation and exploration 
(e.g., Osugi et al. 2005), while in Abernethy et al. (2008), utility balance 
exploits current model knowledge to choose samples with the most uncertain 
prediction.} Such an adaptive query design strategy, based on data from one 
or a few respondents' data, may not be efficient in the early steps of 
sampling. We therefore use an SVM mix for adaptive query design with modest 
computational cost for shrinkage, which samples more efficiently despite 
insufficiency of individual response data, as explained in detail in Section 
3. Our approach is, of course, tailored to product form optimization and the 
trade-off between form and functional attributes, while Huang and Luo (2016) 
method was not designed to address product aesthetics or geometry.

\section{Proposed Model}
\label{subsec:mylabel6}
\subsection{Overview}
\label{subsubsec:mylabel1}
Here we propose and develop a new method aimed at adaptively measuring the 
``utility'' associated with both design elements and traditional product 
(conjoint) attributes. At the heart of the method are iterative ``bi-level'' 
questions. A bi-level question consists of two sequential sub-questions, as 
shown in Figure 1a: one for form alone, the other for both form and 
function. Before delving into specifics of implementation, we must stress 
that they are exactly that: details that enable the method to work quickly 
and reliably with real subjects. The formal properties of the responses used 
here have been chosen to be amenable to scalability, for ease of respondent 
use, and due to the availability of well-vetted algorithms, and \textit{not }because the 
method would not ``work'' with other scale types. That is, although the 
bi-level adaptive questioning method is novel, the overall set-up is 
`modular' in the sense of being able to swap in other question types, as 
computational power allows, although their efficiency, scalability, and 
speed would need careful testing. Rather, the key goal, enabled by the 
bi-level query, is allowing respondents to seamlessly interact with a 
parametric embedding of the product's topology, to achieve real-time visual 
design optimization. 
\begin{figure}
    \centering
    \includegraphics[width=1.0\textwidth]{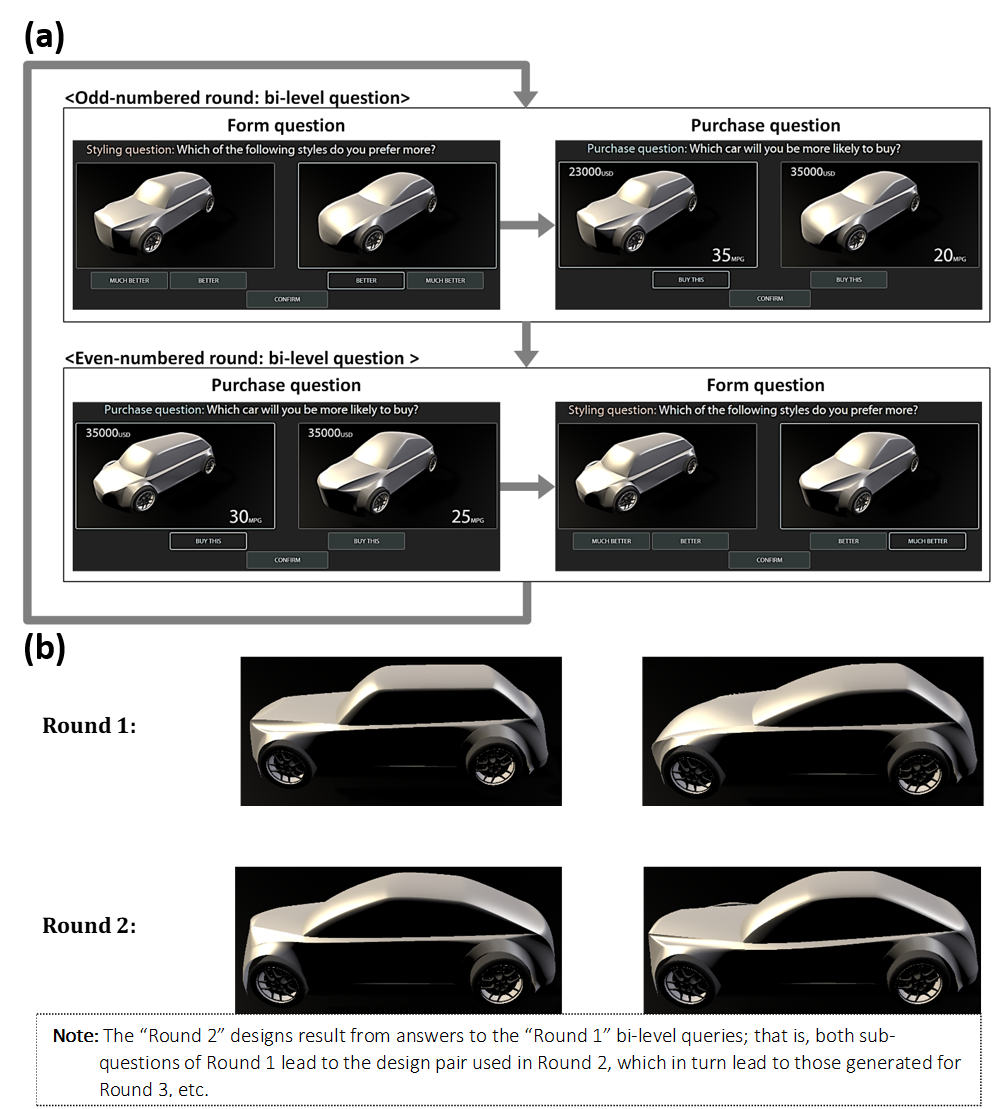}
    \caption{Iterative bi-level Queries and Design Changes (a) Iterative bi-level questions (b) Design query update resulting from prior round choice}
    \label{fig:fig1}
\end{figure}

That said, for the form question, we utilize (as per Figure 1a) a standard 
anchored scale task. Specifically, we present two 3D vehicle renderings and 
ask ``Which of the following styles do you prefer?'' Responses are indicated 
on an ordered 4-point scale: ``left one is much better,'' ``left one is 
better,'' ``right one is better,'' or ``right one is much better.'' Four 
points were used to allow a moderate degree of preference expression over a 
binary choice task, but without exact indifference, which would provide 
little `traction' for the forthcoming adaptive algorithm. Next, for the 
purchase question, we presented the previous 3D vehicle renderings again 
with ``functional attributes,'' such as price and MPG. The respondent was 
then asked ``Which car would you be more likely to buy?'' and made a binary 
choice between the two presented vehicles. Such bi-level questions are 
repeated a specific maximum number of times, set by the analyst. The 
potential tendency for respondents to maintain their choice on the form 
question for the purchase question, irrespective of the newly supplied 
functional attribute information, was controlled for by counterbalancing, 
that is, by switching the order of the two sub-questions from round to 
round, as indicated in Figure 1a.\footnote{ The actual interactive interface 
used for this study can be accessed at 
vehiclechoicemodel.appspot.com. Identifying information required by IRB 
has been removed for journal review.} After a choice was made by the 
respondent, the two designs were updated for the next round of 
(maximally-informative) comparisons, as shown in Figure 1b.

\subsection{Scoring, Utility, and Updating Algorithm}
Design or form preference of individual $i$ stems from the \textit{form model}:
\begin{equation}
\label{eq1}
s_{i}\mathrm{\, =\, }S_{i}\left( \mathrm{\mathbf{x}} \right)\mathrm{\, +\, 
}\mathrm{\varepsilon }_{si}
\end{equation}
where $s_{i}$ is the form score, $S_{i}$ is a non-linear preference 
function, $\mathrm{\mathbf{x\, }}$is a vector of design features 
representing the form, and $\mathrm{\varepsilon }_{si}\mathrm{\, }$is an 
error process. Based on the form score, the overall preference for 
individual $i$ is then given by the following (linear-in-parameters) 
utility model:
\begin{equation}
\label{eq2}
Y\, =\, U_{i}\left( s_{i}\mathrm{,\mathbf{a}} \right)\mathrm{\, =\, 
}\mathrm{\lambda }_{i}s_{i}\mathrm{\, +\, }\mathrm{\beta 
}_{i}^{T}\mathrm{\mathbf{a\, }+\, }\mathrm{\varepsilon }_{yi}
\end{equation}
The vector $\mathrm{\mathbf{a\, }}$consists of binary dummy variables for 
function attribute levels (i.e., a three-entry binary vector for a 
four-leveled attribute); $\mathrm{\lambda }_{i}$ is the weight of the 
form score; $\mathrm{\beta }_{i}$ is the partworth vector for functional 
attribute levels; and $\mathrm{\varepsilon }_{yi}\mathrm{\, }$is 
associated error. In other words, Eq. (\ref{eq2}) is a standard conjoint utility 
specification, including the form score $(s_{i})$, to 
be calibrated via Eq. (\ref{eq1}) and weighted via $\mathrm{\lambda 
}_{i}\mathrm{,\, }$using Eq. (\ref{eq2}). [Note that it is further possible to 
include interaction terms in the specification for $U_{i}\left( 
s_{i}\mathrm{,\mathbf{a}} \right)$, for example, between form ($s_{i})$ and 
various attributes within $\mathrm{\mathbf{a}}$, although this can lead to 
many additional estimated parameters and identification difficulties. 
Because, as shown later, doing so did not improve accuracy in our 
application we do not raise this possibility explicitly again, referring 
generically to $U_{i}\left( s_{i}\mathrm{,\mathbf{a}} \right)$ as 
linear-in-parameters.]

The two preference models, Eq. (\ref{eq1}) and (\ref{eq2}), are updated iteratively in real 
time by bi-level questioning, as per Figure 1a. The process unfolds as 
follows:
{\setstretch{1.08}
\begin{itemize}
\item \textbf{Form question}: an individual makes a metric paired-comparison between two forms created by design variables, $\mathrm{\mathbf{x}}^{\mathrm{(1)}}$ and $\mathrm{\mathbf{x}}^{\mathrm{(2)}}$. The preference model $S_{i}\left( \mathrm{\mathbf{x}} \right)$ in Eq. (\ref{eq1}) is trained; then form scores, $s_{i}^{\mathrm{1)}}$ and $s_{i}^{\mathrm{(2)}}$, are estimated. Finally, two function attributes (price and MPG in our application), $\mathrm{\mathbf{a}}^{\mathrm{(\ref{eq1})}}$ and $\mathrm{\mathbf{a}}^{\mathrm{(2)}}$, are sampled for the subsequent purchase question.
\item \textbf{Purchase question}: an individual makes a binary choice between two bundles of form and functions  $[s_{i}^{\left( \mathrm{1} \right)}\mathrm{\mathbf{a}}^{\left( \mathrm{1} \right)}]$  and  $\mathrm{[}s_{i}^{\mathrm{(2)}}\mathrm{\mathbf{a}}^{\mathrm{(2)}}\mathrm{]}$. The weight of the form score $\mathrm{\lambda }_{i}$ and the partworths for functions $\mathrm{\beta }_{i}$ in Eq. (\ref{eq2}) are estimated. Finally, two forms,  $\mathrm{\mathbf{x}}^{\mathrm{(1)}}$ and $\mathrm{\mathbf{x}}^{\mathrm{(2)}}$, were sampled for the subsequent form question.
\end{itemize}
}
The overall process for querying, sampling, and learning is summarized 
visually, in flow chart form, in Figure 2, and verbally as follows:
\begin{figure}
    \centering
    \includegraphics[width=1.0\textwidth]{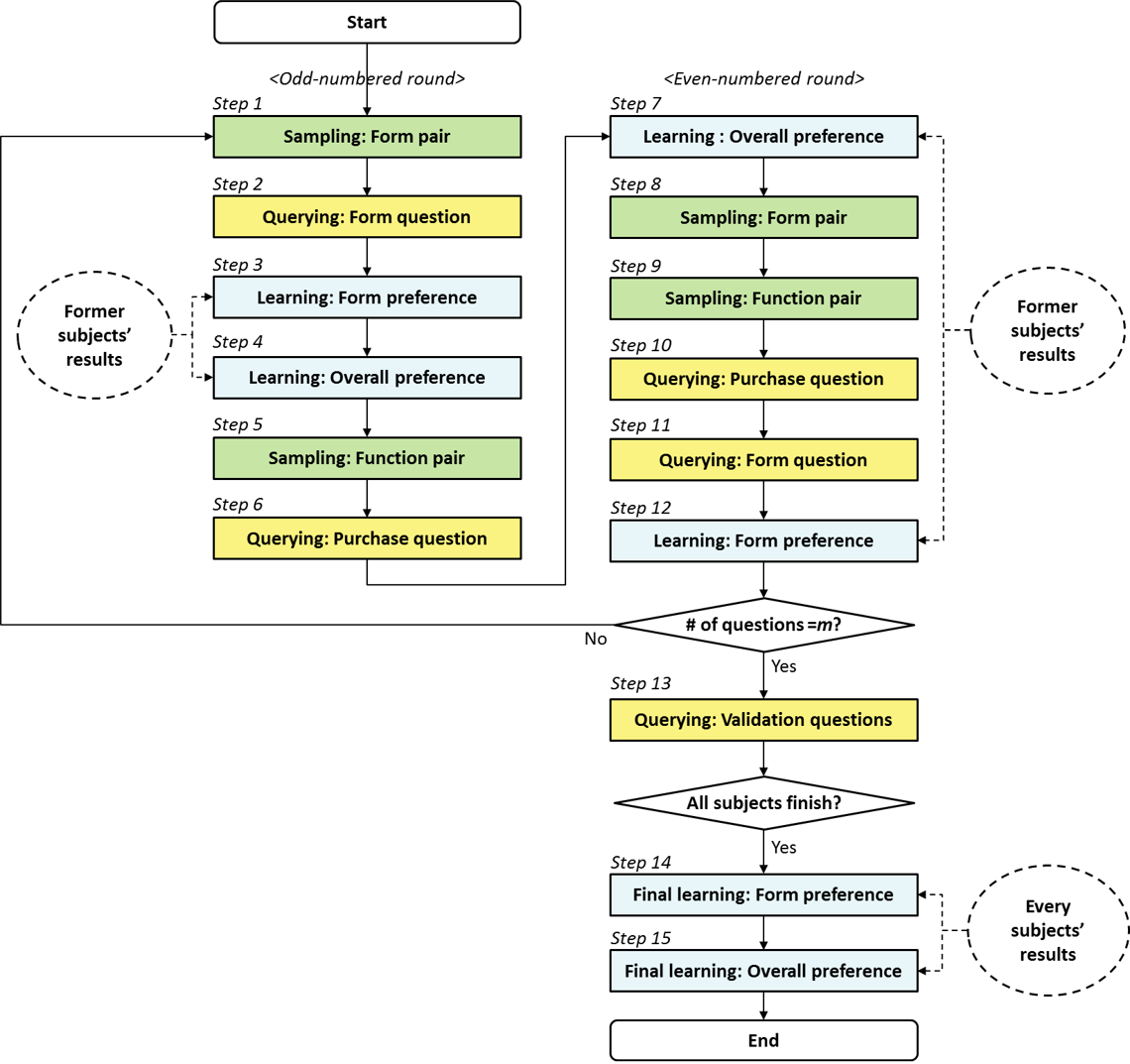}
    \caption{Overall process for querying, sampling, and learning }
    \label{fig:fig2}
\end{figure}

{\setstretch{1.0}
\begin{itemize}
\item \textbf{Start}. A new questionnaire is initialized when an individual accesses the website.
\item \textbf{Step 1}: \textit{Sampling form pair}. A pair of vehicle renderings is generated from the design space based on the current form preference model. The pair is such that their expected form \textit{preference} is roughly equal, but their \textit{shapes} differ maximally from one another and from all forms used before. If this is the first question for the current subject, a fixed form pair is used.
\item \textbf{Step 2}: \textit{Querying form question}. A metric paired-comparison response is received from the subject.
\item \textbf{Step 3}: \textit{Learning form preference}. A form preference model is trained based on previous form responses from this subject. Once the form model is learned (or updated if not the first round), the \textit{form scores} of previously sampled vehicle renderings for the current subject are updated. Former subjects' form preference models are used for shrinkage.
\item \textbf{Step 4}: \textit{Learning overall preference}. Except for the first round, the overall preference model was adjusted based on the updated form scores. Former subjects' overall preference models were used for shrinkage.
\item \textbf{Step 5}: \textit{Sampling function pair}. Generate function attributes (e.g., price and MPG) for the current pair of vehicle forms, based on the updated overall preference model.
\item \textbf{Step 6}: \textit{Querying purchase question}. A (binary) choice is received from the subject once the function attributes are shown along with the forms.
\item \textbf{Step 7}: \textit{Learning overall preference}. Same as Step 4. [Steps 1 - 7 complete an odd-numbered round. Even-numbered rounds switch the order of the form and purchase questions, as elaborated in steps 8-12 below.]
\item \textbf{Steps 8-12}: Same as Steps 1 (\textit{Sampling form pair}); 5 (\textit{Sampling function pair}); 6 (\textit{Querying purchase question}); 2 (\textit{Querying form question}); and 3 (\textit{Learning form preference}). [If the iteration has reached its maximum round number, $m$, go to Step 13; otherwise, to 1.]
\item \textbf{Step 13}: \textit{Querying validation question}. Several validation sets are presented and used to check hit rate. [If all subjects have finished, go to Step 14; otherwise, wait.]
\item \textbf{Step 14}: \textit{Final learning form preference}. Finalize individual-level form preference models using all other subjects' results.
\item \textbf{Step 15}: \textit{Final learning overall preference}. Finalize individual-level overall preference models using all other subjects' results.
\item \textbf{End}. Hit rate checked using responses to the validation questions.
\end{itemize}
} 
Because the purchase question can be characterized as ``forced binary 
choice'', a natural question concerns whether this preferable to, for 
example, a three-option query that includes ``no choice'' or an ``outside 
option''. While this is possible, a pilot pretest suggested that some 
participants (recruited through crowdsourcing) briskly dispensed with the 
survey by frequently selecting ``no choice'', although this may occur less 
with highly committed or remunerated participants. We also direct the reader 
to Brazell et al. (2006) for additional advantages, e.g., disentangling 
``which'' option(s) consumers prefer from volumetric predictions regarding 
``whether'' they will purchase at all. [Appendix B provides amendments to 
the ``no choice'' set-up, which entail no changes in core implementation 
algorithms.] 

During each survey, we use the rank SVM mix algorithm (Evgeniou et al. 2005) 
for rapid training and an uncertainty sampling scheme similar to Settles 
(2010) and Tong and Koller (2002) for real-time generation of comparison 
pairs. We follow the suggestion of Toubia et al. (2013), ``Once the data 
have been collected, we recommend combining our adaptive questionnaire 
design method with hierarchical Bayes, a well-proven method for estimating 
the parameters given the data''. That is, after all user data are collected, 
we estimate individual partworths using standard hierarchical Bayesian (HB) 
techniques (Lenk et al. 1996, Rossi and Allenby 2003). We note that, unlike 
HB, SVM does not rely on an explicit notion of likelihood, although we show 
in Appendix C that it is consistent with hinge-loss and a Gaussian prior 
(similar to Evgeniou, Pontil, and Toubia 2007, p. 807). We emphasize again 
that the overarching method is agnostic on specific (machine learning) 
algorithms, which can be replaced with alternatives suited to the analyst's 
specific survey environment and research goals, dependent on computational 
speed and scalability (a topic we explore later empirically). 

We next elaborate on how these algorithms are applied: Section 3.3 discusses 
learning methods for both the form and overall preference models, as well as 
the rationale for handing these separately, while Section 3.4 addresses 
sampling methods to generate pairs for comparison.

\subsection{Learning Preferences}
\label{subsubsec:mylabel2}
\paragraph{3.3.1 Form Preference Learning}
\label{para:mylabel1}
As mentioned earlier, the form preference model in Eq. (\ref{eq1}) is trained using 
a rank SVM mix algorithm. The idea is to fit a model consistent with the 
metric paired-comparison results, i.e., if one form of the pair is much 
preferred to the other, the preference gap between the two should also be 
larger than a pair that is less differentiated, i.e., one is preferred to 
the other. Following the treatments in Joachims (2002) and Chapelle and 
Keerthi (2010), the training problem can be formulated as follows:
\begin{equation}
\label{eq3}
\begin{array}{ll}
\mathop{\mathrm{min}}_{\boldsymbol{\mathrm{w}}} & \mathrm{\ \ \ }{\boldsymbol{\mathrm{w}}}^T\boldsymbol{\mathrm{w}} \\ 
\mathrm{subject\ to} & {\mathrm{\ \ \ }\boldsymbol{\mathrm{w}}}^T\mathrm{\phi }\mathrm{(}{\boldsymbol{\mathrm{x}}}^{\mathrm{(1)}}_j\mathrm{)} \; - \;  {\boldsymbol{\mathrm{w}}}^T\mathrm{\phi }\mathrm{(}{\boldsymbol{\mathrm{x}}}^{\mathrm{(2)}}_j\mathrm{)} \; \mathrm{\ge } \; c_j\mathrm{,\ }\mathrm{\forall }j\mathrm{=1,\dots ,}m \\ 
\mathrm{where} & \mathrm{\ \ \ }c_j \; \mathrm{\in } \; \mathrm{\{}\mathrm{1,2}\mathrm{\}}\mathrm{.} \end{array}
\end{equation}
Here  $\mathrm{\mathbf{x}}_{j}^{\mathrm{(1)}}$ and $
\mathrm{\mathbf{x}}_{j}^{\mathrm{(2)}}$ are the design features for the 
chosen and unchosen forms in the $j$-th questions, respectively. User 
responses are represented by $c_{j}$: when $
\mathrm{\mathbf{x}}_{j}^{\mathrm{(1)}}$ is ``better'' than $
\mathrm{\mathbf{x}}_{j}^{\mathrm{(2)}}$, $c_{j}$ is set to $
\mathrm{1}$; when $\mathrm{\mathbf{x}}_{j}^{\mathrm{(1)}}$ is ``much 
better'', $c_{j}$ is set to $\mathrm{2}$. The objective in (\ref{eq3}) 
follows a hard-margin SVM formulation where $
\mathrm{\mathbf{w}}^{T}\mathrm{\mathbf{w}}\mathbf{\, }$represents the model 
complexity. Later, in our simulation studies, we will explore both the 
sensitivity of the hard-margin SVM to choices of $c_{j}$, as well as its 
``brittleness'' to contradictory preference ordering on the part of 
respondents. It is possible to view (\ref{eq3}) in terms of its Lagrangian, $L\left( 
\mathrm{\mathbf{w,\alpha }} \right)$; Appendix C details its relation to a 
negative log-likelihood with parameters $\mathrm{\mathbf{\alpha }}\, 
$balancing the Gaussian prior ($\mathrm{\mathbf{w}}\sim 
N(\mathbf{0}, \mathrm{\mathbf{I}}))$ and the data (i.e., 
$\mathrm{\mathbf{x}}_{j}^{\left( 1 \right)}$ preferred to 
$\mathrm{\mathbf{x}}_{j}^{\left( 2 \right)})$.

One can project $\mathrm{\mathbf{x\, }}$ to a high-dimensional space, 
i.e., $\mathrm{\mathbf{x}\to \phi (\mathbf{x})}$, so that the constraints 
for $\mathrm{\forall }j\mathrm{=1,\mathellipsis ,}m$  can be satisfied. 
The dual problem of (\ref{eq3}) can be expressed as follows:
\begin{equation}\label{eq4}
 \begin{array}{ll}
\mathop{\mathrm{min}}_{\mathrm{\alphaup }} & \frac{\mathrm{1}}{\mathrm{2}}{\boldsymbol{\mathrm{\alphaup }}}^T\boldsymbol{\mathrm{Q}}\boldsymbol{\mathrm{\alphaup }} \; \mathrm{-} \; {\boldsymbol{\mathrm{c}}}^T\boldsymbol{\mathrm{\alphaup }} \\ 
\mathrm{subject\ to} & \boldsymbol{\mathrm{\alphaup }} \; \mathrm{\ge } \; \mathrm{0,} \end{array}
\end{equation}
where $\mathrm{\mathbf{\alpha }\, }$are Lagrangian multipliers and $
\mathrm{\mathbf{Q}}\mathbf{\, }$is an $m$ by $m$ matrix with each 
element$,\, Q_{ij},$ being the inner product  $
\left\langle \mathrm{\phi }\mathrm{(}{\boldsymbol{\mathrm{x}}}_i\mathrm{),}\mathrm{\phi }\mathrm{(}{\boldsymbol{\mathrm{x}}}_j\mathrm{)}\right\rangle
$. Following Chang and Lin (2011), a common choice for this inner product 
relies on the Gaussian kernel, which we denote as $K$:
\begin{equation}
\label{eq5}
K\left({\boldsymbol{\mathrm{x}}}_i,{\boldsymbol{\mathrm{x}}}_j\right) \; \mathrm{\equiv } \; \left\langle \mathrm{\phi }\mathrm{(}{\boldsymbol{\mathrm{x}}}_i\mathrm{),}\mathrm{\phi }\mathrm{(}{\boldsymbol{\mathrm{x}}}_j\mathrm{)}\right\rangle \;  \mathrm{:=exp(-} \mathrm{\gammaup }{\left\|{\boldsymbol{\mathrm{x}}}_i\mathrm{-}{\boldsymbol{\mathrm{x}}}_j\right\|}^{\mathrm{2}}\mathrm{),} 
\end{equation}
where the Gaussian parameter $\mathrm{\gamma }$ is set at $\mathrm{\gamma 
=1/(number\, of\, design\, features)}$.\footnote{ Note that this does not 
uniquely specify $\phi $, only the ``inner product'' function. In fact, 
defining $\phi $ is unnecessary here, since the dual problem to (\ref{eq4}), 
owing to ``strong duality'', only requires the definition of the inner 
product $\phi^{T}\phi $ for the calculation of matrix Q.} The dual problem 
can then be solved efficiently following the algorithm of Fan et al. (2005): 
based on the resultant Lagrangian multipliers, $\mathrm{\mathbf{\alpha 
}}$, user preferences on a given form  $\mathrm{\mathbf{x\, }}$ can be 
quantified as
\begin{equation}
\label{eq6}
S\mathrm{(\mathbf{x})=}\sum\limits_{j\mathrm{=1}}^m  \; {\mathrm{\alpha 
}_{j}\mathrm{(}K\mathrm{(\mathbf{x},}\mathrm{\mathbf{x}}_{j}^{\mathrm{(1)}}\mathrm{)} \; - \; K\mathrm{(\mathbf{x},}\mathrm{\mathbf{x}}_{j}^{\mathrm{(2)}}\mathrm{))}} \end{equation}
where  $\mathrm{\mathbf{x}}_{j}^{\mathrm{(1)}}$ are all chosen forms 
during the survey and  $\mathrm{\mathbf{x}}_{j}^{\mathrm{(2)}}$ the 
unchosen ones. For stability and comparability, the design features  $
\mathrm{\mathbf{x\, }}$ are normalized to have zero mean and unit standard 
deviation before being used in training and prediction. Note that a 
soft-margin SVM formulation (Cortes and Vapnik 1995, Cristianini and 
Shawe-Taylor 2000) could be used in place of (\ref{eq4}) to deal with `noisy' user 
responses, a topic we examine later via simulation for the hard-margin SVM.

Due to the limited data collection from individuals, it is desirable to 
leverage data collected from \textit{all} participants to improve the robustness and 
accuracy of individual-level preference models, similar to the shrinkage 
underlying HB methods. As discussed earlier, we use the (modest 
computational cost) method of Evgeniou et al. (2005), with partworth of 
participant $i$ given as
\begin{equation}
\label{eq7}
\mathrm{\mathbf{w\, }=\, }\sum\nolimits_j {} \mathrm{\alpha }_{j}\left( 
\mathrm{\phi (}\mathrm{\mathbf{x}}_{j}^{\mathrm{(1)}}\mathrm{)-\phi 
(}\mathrm{\mathbf{x}}_{j}^{\mathrm{(2)}}\mathrm{)} \right),
\end{equation}
and population-wise partworth as
\begin{equation}
\label{eq8}
\bar{w}\mathrm{\, =\, }\frac{\mathrm{1}}{N}\sum\nolimits_{n\mathrm{=1}}^N {} 
\sum\nolimits_j {} \mathrm{\alpha }_{j}^{\mathrm{(}n\mathrm{)}}\left( 
\mathrm{\phi 
(}\mathrm{\mathbf{x}}_{j}^{\mathrm{(1,}n\mathrm{)}}\mathrm{)-\phi 
(}\mathrm{\mathbf{x}}_{j}^{\mathrm{(2,}n\mathrm{)}}\mathrm{)} \right)
\end{equation}
where $N$ is the total number of participants, $\mathrm{\alpha 
}_{j}^{\mathrm{(}n\mathrm{)}}$ is the Lagrangian multiplier for the  $
j$-th question for participant  $n$, and  $
\mathrm{\mathbf{x}}_{j}^{\mathrm{(1,}n\mathrm{)}}$ and $
\mathrm{\mathbf{x}}_{j}^{\mathrm{(2,}n\mathrm{)}}$ are the chosen and 
unchosen forms in that question, respectively. With a weighting factor $
\mathrm{\eta }_{i}\mathrm{\in [0,1]}$, the individual form preference for 
participant $i$ and a given form $\mathrm{\mathbf{x\, }}$can be 
calculated as:
\begin{equation}
\label{eq9}
 \begin{array}{l}
S^{\mathrm{*}}_i\left(\boldsymbol{\mathrm{x}}\right)\mathrm{\ =\ (}{\mathrm{\etaup }}_i\boldsymbol{\mathrm{w}}\mathrm{+(1-}{\mathrm{\etaup }}_i\mathrm{)}\overline{\boldsymbol{w}}{\mathrm{)}}^T\mathrm{\phi }\mathrm{(}\boldsymbol{\mathrm{x}}\mathrm{)} \\ 
\ \ \ \ \ \ \ \ \ \mathrm{\ =}{\mathrm{\ }\mathrm{\etaup }}_i\sum_j{}{\mathrm{\alphaup }}_j\left(\left\langle \mathrm{\phi }\mathrm{(}{\boldsymbol{\mathrm{x}}}^{\mathrm{(1)}}_j\mathrm{),}\mathrm{\phi }\mathrm{(}\boldsymbol{\mathrm{x}}\mathrm{)}\right\rangle \mathrm{-}\left\langle \mathrm{\phi }\mathrm{(}{\boldsymbol{\mathrm{x}}}^{\mathrm{(2)}}_j\mathrm{),}\mathrm{\phi }\mathrm{(}\boldsymbol{\mathrm{x}}\mathrm{)}\right\rangle \right) \\ 
\ \ \ \ \ \ \mathrm{\ \ \ \ \ \ \ \ \ \ \ \ +\ (1-}{\mathrm{\etaup }}_i\mathrm{)}\frac{\mathrm{1}}{N}\sum^N_{n\mathrm{=1}}{}\sum_j{}{\mathrm{\alphaup }}^{\mathrm{(}n\mathrm{)}}_j\left(\left\langle \mathrm{\phi }\mathrm{(}{\boldsymbol{\mathrm{x}}}^{\mathrm{(1,}n\mathrm{)}}_j\mathrm{),}\mathrm{\phi }\mathrm{(}\boldsymbol{\mathrm{x}}\mathrm{)}\right\rangle \mathrm{-}\left\langle \mathrm{\phi }\mathrm{(}{\boldsymbol{\mathrm{x}}}^{\mathrm{(2,}n\mathrm{)}}_j\mathrm{),}\mathrm{\phi }\mathrm{(}\boldsymbol{\mathrm{x}}\mathrm{)}\right\rangle \right) \\ 
\ \ \ \ \ \ \ \ \ \mathrm{\ =\ }{\mathrm{\etaup }}_iS_i\mathrm{(}\boldsymbol{\mathrm{x}}\mathrm{)+(1-}{\mathrm{\etaup }}_i\mathrm{)}\frac{\mathrm{1}}{N}\sum^N_{i\mathrm{=1}}{}S_i\mathrm{(}\boldsymbol{\mathrm{x}}\mathrm{)} \end{array}
\end{equation}
We note that Eq. (\ref{eq9}) is used only for the adaptive query design, and not for 
coefficient estimation (which is fully Bayesian), so values of $\mathrm{\eta 
}_{i}$ primarily affect generation efficiency. Specifically, if $
\mathrm{\eta }_{i}$ is small, the function of individual $i$ shrinks 
strongly toward the population-level function. For active learning during 
the survey, $\mathrm{\eta }_{i}$ can be selected at the discretion of 
the analyst: when there are few prior respondents, a large value of $
\mathrm{\eta }_{i}$ can be used; otherwise, a smaller $\mathrm{\eta 
}_{i}$ value is appropriate. For estimation of form preference after 
finishing the survey, optimal $\eta_{i}$ for the final estimation can be 
determined by cross-validation. That is, when we use $J$ questions, 
$J\mathrm{-1}$ responses are used for training and the remaining response is 
used for assessing prediction performance. $J$ rounds of cross-validation 
are performed with different test data to minimize prediction error. In our 
experiments, we used $\mathrm{\eta }_{\mathrm{1}}=1$ and $\mathrm{\eta 
}_{N}=0.7$ for the first respondent and the last respondent, respectively, 
with intermediate respondents interpolated linearly between these values.

\paragraph{3.3.2 Overall Preference Learning}
\label{para:mylabel2}
The coefficients in the overall preference model of Eq. (\ref{eq2}) can be estimated 
analogously to those of the form preference model, that is, using a rank SVM 
mix algorithm for active learning during surveys and HB for population-level 
modeling. The problem formulation for individual-level learning is as 
follows:
\begin{equation}
\label{eq10}
\begin{array}{ll}
\mathrm{\ }\mathop{\mathrm{min}}_{{\boldsymbol{\mathrm{W}}}_i} & {\boldsymbol{\mathrm{W}}}^T_i{\boldsymbol{\mathrm{W}}}_i \\ 
\mathrm{subject\ to} & {\boldsymbol{\mathrm{W}}}^T_i{\boldsymbol{\mathrm{X}}}^{\mathrm{(1)}}_{ij} \;  \mathrm{-} \; {\boldsymbol{\mathrm{W}}}^T_i{\boldsymbol{\mathrm{X}}}^{\mathrm{(2)}}_{ij} \; \mathrm{\ge } \; \mathrm{1,\ }\mathrm{\forall }j\mathrm{=1,\dots ,}m, \end{array}
\end{equation}
where  $\mathrm{\mathbf{W}}_{i}^{T} = \left[ \mathrm{\lambda 
}_{i}, \mathrm{\beta }_{i}^{T} \right]$  are the linear 
coefficients, and  $\mathrm{\mathbf{X}}_{ij}^{T} = \left[ 
s_{ij}, \mathrm{\mathbf{a}}_{ij}^{T} \right]$  consists of the form 
score and the binary dummy variables of the function attributes for the $
j$-th comparison for individual $i$. Just as for (\ref{eq4}), we apply the Fan et al. (2005) algorithm to the dual of (\ref{eq10}). All constraints are set to be 
greater than or equal to 1, as the comparison in this case is binary rather 
than metric. In Appendix D, we demonstrate that ``1'' can be changed 
arbitrarily to any positive value, and also write out the dual of (\ref{eq10}) 
explicitly. This dual is solved using a linear mapping, i.e.,  $
\mathrm{\phi (}\mathrm{\mathbf{X}}_{ij}\mathrm{)=}\mathrm{\mathbf{X}}_{ij}$, 
and the resultant individual-wise partworth vector  $
\mathrm{\mathbf{W}}_{i}$ can be expressed as:
\begin{equation}
\label{eq11}
\mathrm{\mathbf{W}}_{i} = \sum\nolimits_{j\mathrm{=1}}^m {} 
\mathrm{(}\mathrm{\mathbf{X}}_{ij}^{\mathrm{(1)}} - \mathrm{\mathbf{X}}_{ij}^{\mathrm{(2)}}\mathrm{)}^{T}\mathrm{\alpha 
}_{ij}
\end{equation}
In order to leverage population-level data, the linear shrinkage method of 
Evgeniou et al. (2005) is again used for individual-level partworths, $
\mathrm{\mathbf{W}}_{i}^{\mathrm{\ast }}$; specifically,
\begin{equation}
\label{eq12}
\mathrm{\mathbf{W}}_{i}^{\mathrm{\ast }} = \mathrm{\eta 
}_{i}\mathrm{\mathbf{W}}_{i} + \mathrm{(1} - \mathrm{\eta 
}_{i}\mathrm{)}\frac{\mathrm{1}}{N}\sum\nolimits_{i\mathrm{=1}}^N {} 
\mathrm{\mathbf{W}}_{i}
\end{equation}
where $\mathrm{\eta }_{i}$ is the weight for individual $i$, and $
N$ is the number of individuals.\footnote{ Shrinkage weights are constant 
across preference vector elements, so do not leverage possible covariances 
across them. Fixed SVM shrinkage is appropriate when finding the optimal 
variance-covariance matrix is not affordable for real-time query generation, 
and our operationalization is consistent with the standard LIBSVM package. 
Note that, for a linear function, partworth covariances can be optimized 
along with means; but for a nonlinear function that maps samples to an 
infinite-dimensional space, this is infeasible.} For population-level 
preference modeling, a hierarchical binary logit model (Rossi et al. 2005), 
with weakly-informative and zero-centered priors, is used for estimation. 
Specifically, at the upper level of the Bayesian model, we assume $
\mathrm{\mathbf{W}}_{i}$ to have a multivariate normal distribution,
\begin{equation}
\label{eq13}
\mathrm{\mathbf{W}}_{i}\mathrm{\, \sim \, }N\left( 
\mathrm{\mathbf{0},\Lambda } \right)
\end{equation}
At the lower level, choice probabilities take binary logit form:
\begin{equation}
\label{eq14}
\mathrm{Pr}\left( y_{ij} = 1 \right) = \mathrm{(1+exp[}\mathrm{\mathbf{W}}_{i}^{T}
\mathrm{(}\mathrm{\mathbf{X}}_{ij}^{\mathrm{(2)}}
 - \mathrm{\mathbf{X}}_{ij}^{\mathrm{(1)}}\mathrm{)]}\mathrm{)}^{\mathrm{-1}}
\end{equation}
where  $\mathrm{Pr}\left( y_{ij} = 1 \right)$  and  $
\mathrm{Pr}\left( y_{ij} =0 \right)$  denote the probabilities of 
selecting  $\mathrm{\mathbf{X}}_{ij}^{\mathrm{(1)}}$  and  $
\mathrm{\mathbf{X}}_{ij}^{\mathrm{(2)}}$, respectively, for the  $j$-th 
question of individual  $i$. 

\paragraph{3.3.3 Differences in Learning Algorithms and Rationale for Separation}
It is important to note that the formulations are different for form 
preference and for overall utility: while the latter is governed by an 
identity feature function, for the former, a Gaussian kernel is applied to 
the geometric features of each form, where the geometric feature vector 
contains all pairwise distances from the control points that define the 3D 
geometries; these control points are further governed by the input variables 
in a nonlinear way. Due to the infinite-dimensional nature of the feature 
space induced by the Gaussian kernel, learning or modeling the covariance 
matrix of the partworths during querying is infeasible, and is an identity 
for shrinkage purposes. The ``online'' estimation of SVM model parameters is 
thereby based on both current and previous responses: rather than 
formulating and solving a large SVM problem with all responses considered, 
for computational feasibility, parameter estimates from each individual 
survey are weighted linearly, with earlier responses receiving lower 
weights. That is, a speedy heuristic approach is used during the survey and 
query generation, and a formal, computationally costly Bayes model for 
subsequent (offline) estimation. 

\subsection{Sampling Questions}
\label{subsubsec:mylabel3}
We adaptively sample the next pair of forms or functional attributes based 
on two criteria. First, a profile pair comprising a question should have as 
near to the same utility as possible according to the current model. The 
second criterion is to maximize the minimum distance from existing data 
points, from both the current participant and all previous ones. The 
implementation is as follows: among the pair, the first form (or function 
attribute set) is sampled solely by the second criterion:
\begin{equation}
\label{eq15}
 \begin{array}{ll}
\mathop{\mathrm{max}}_{{\boldsymbol{\mathrm{x}}}^{new}_{\mathrm{1}}\mathrm{\in }{\left[\mathrm{0,1}\right]}^{\mathrm{19}}} & {\mathop{\mathrm{min}}_{j} {\left\|{\boldsymbol{\mathrm{x}}}^{new}_{\mathrm{1}}\mathrm{-}{\boldsymbol{\mathrm{x}}}^{old}_j\right\|}^{\mathrm{2}}\ } \\ 
\mathrm{subject\ to} & lb \; \mathrm{\le } \; {\boldsymbol{\mathrm{x}}}^{new}_{\mathrm{1}} \; \mathrm{\le }  \; ub, \end{array}
\end{equation}
where $\mathrm{\mathbf{x}}_{\mathrm{1}}^{new}$ is the first form (or 
function attribute set) alternative for the next question, $
\mathrm{\mathbf{x}}_{j}^{old}$ are the $j$-th form alternatives used in 
previous questions, and $m_{\mathrm{\mathbf{x}}^{old}}$ is the number of 
form alternatives used previously.

Once  $\mathrm{\mathbf{x}}_{\mathrm{1}}^{new}$  is sampled, its form 
preference value (or utility) can be calculated based on the current model. 
The second sample, $\mathrm{\mathbf{x}}_{\mathrm{2}}^{new}$, is obtained 
via optimizing a weighted sum of the two criteria:
\begin{equation}
\label{eq16}
 \begin{array}{ll}
\mathrm{\ }\mathop{\mathrm{min}}_{{\boldsymbol{\mathrm{x}}}^{new}_{\mathrm{2}}\mathrm{\in }{\left[\mathrm{0,1}\right]}^{\mathrm{19}}} & {v_{\mathrm{1}}\mathrm{exp} \left(\mathrm{-}{\left\|S\left({\boldsymbol{\mathrm{x}}}^{new}_{\mathrm{1}}\right) - S\left({\boldsymbol{\mathrm{x}}}^{new}_{\mathrm{2}}\right)\right\|}^{\mathrm{2}}\right) + v_{\mathrm{2}}\mathrm{(}{\left\|{\boldsymbol{\mathrm{x}}}^{new}_{\mathrm{1}} - {\boldsymbol{\mathrm{x}}}^{new}_{\mathrm{2}}\right\|}^{\mathrm{2}} + {\mathop{\mathrm{min}}_{j} {\left\|{\boldsymbol{\mathrm{x}}}^{new}_{\mathrm{2}} - {\boldsymbol{\mathrm{x}}}^{old}_j\right\|}^{\mathrm{2}}\ }\mathrm{)}\ } \\ 
\mathrm{subject\ to} & lb \; \mathrm{\le } \; {\boldsymbol{\mathrm{x}}}^{new}_{\mathrm{2}} \; \mathrm{\le } \; ub, \end{array}
\end{equation}
where  $S\left( \mathrm{\mathbf{x}} \right)$ is the form preference model 
and  $v_{\mathrm{1}}$  and  $v_{\mathrm{2}}$  are the weights. (\ref{eq16}) 
again balances two objectives: the preference values of the two new designs 
should be similar, and the second design should differ from the first. By 
construction, $\mathrm{\mathbf{x}}_{\mathrm{2}}^{new}$  should be far away 
from not only previous samples, but also from the current first sample; 
$v_{\mathrm{1}}$  and  $v_{\mathrm{2}}$  (or, equivalently, their ratio) 
are chosen by the experimenter to balance the two criteria (in our 
experiments, $v_{1}=0.99$ and $v_{2}=0.01$; Appendix E provides details on 
setting these values and results of using others). Due to high potential 
nonconvexity, locating each successive form pair is accomplished via genetic 
algorithms (GAs), by enumerating all combinations of attribute levels using 
Eq. (\ref{eq15}) and (\ref{eq16}).\footnote{ Details for the GA implementation are: for Eq. 
(\ref{eq15}), population size $=$ 20, max. generations $=$ 100; for Eq. (\ref{eq16}), 
population size $=$ 50, max. generations $=$ 500. Each iteration generates a 
set of parent pairs using the current population, where parent set size $=$ 
population size. Pair generation is via a tournament scheme: two tournaments 
are played, each with set of (population size - 1) players uniformly 
randomly chosen from the population, and the ``most fit'' chosen as the 
parent. This pair then goes through a one-point crossover procedure, with 
cutting point uniformly randomly chosen: if the two parents are 
$x_{1}=[x_{11},\, x_{12}]$ and $x_{2}=[x_{21},\, x_{22}]$, the two output 
designs from crossover are $[x_{11},\, x_{22}]$ and $[x_{12},\, x_{21}]$, 
with a 0.1 mutation probability. The mutation operation picks one element of 
the vector $x$ at a random location $i$ (uniformly), and changes its value 
from $x_{i}$ to $x_{i}+\delta $, where $\delta $ is uniformly drawn from 
$[-0.05,\, 0.05]$; the mutated value is bounded in $[0,1]$. The algorithm 
terminates when the maximum number of generations is reached.} 

In simple terms, the active learning algorithm generates designs that are 
different from one another (and from existing designs) while being 
similar in preference (according to the current model). These two principles 
serve the purposes of exploitation and exploration, respectively, which are 
commonly enacted in active learning algorithms, e.g., Tong and Koller 
(2001), Osugi et al. 2005), and Abernethy et al. (2008). We note that it is 
possible to optimize both questions in the bi-level query pair 
simultaneously rather than individually in a sequence, as per Eq. (\ref{eq15}) and 
(\ref{eq16}). A pragmatic challenge in doing so is computational speed of the 
application engine (exogenously set to 30 seconds in our Google-based 
implementation), and user tolerance for waiting, which is considerably 
shorter. Sequential optimization / generation of the question pair is 
thereby a pragmatic compromise, not a theoretical limitation. Appendix F 
provides full details for simultaneous (``all-in-one'') pair generation, as 
well as benchmarks relative to the sequential approach that support the 
superiority of the decompositional approach. 

\section{Application: Geometric Set-Up and Model Simulation}
\label{subsec:mylabel7}
We demonstrate the benefits of the proposed approach by applying it to the 
design of a highly multiattribute, visually complex durable: a passenger 
sedan. As discussed in our overview of the literature, vehicle design has 
been among the main application domains of form design optimization models 
in the engineering discipline, and was proposed as a canonical application 
of the presentation of pictorial information in conjoint by Green and 
Srinivasan (1990).

To apply any method for form optimization requires a way to explore the 
space of designs. Both Green and Srinivasan (1990) and Dotson et al. (2019), 
as well as the overwhelming bulk of real-world applications in the marketing 
discipline, rely on a candidate set of images, which as discussed previously 
make interpolation or extrapolation precarious, along with parsimony 
challenges for heterogeneity estimation. An alternative approach common in 
engineering and some prior marketing research (e.g., Michalek et al. 2005) 
is an explicit product topology model, that is, a geometric representation 
of the external form and internal workings of the product. Here, our goal is 
less onerous, as form optimization only requires parameterizing the external 
(3D) shape of the product in question, not ensuring that, for example, it is 
possible to engineer \textit{internal} components to conform with cost and safety 
constraints.

For vehicle form representation, we therefore developed a 3D parametric 
vehicle shape model (Ren and Papalambros 2011 provide additional technical 
detail; see as well Orsborn et al. 2009). This parametric model generates 3D 
renderings using two-leveled structures, as follows. First, nineteen design 
variables $\mathrm{\mathbf{x}}\mathrm{\mathbf{\, }}$(ranging from 0 to 
1; see Figure 3) were determined to be sufficient to realistically set the 
coordinates of \textit{control points}, which in turn generate (Bezier) surfaces of the 3D model. 
Examples of the underlying parameterization include the distance from the 
front grille to the midpoint of the hood, the elevation and pitch of the 
center of the windshield's highest point, etc. The full space of potential 
sedan shapes would doubtless require additional parameters -- for example, 
door and window shape are not explicitly optimized -- but the 19 variables 
used here provide a very wide array of configurations, covering a broad 
swath of tested models currently in the North American sedan market, and 
thereby provide a reasonable trade-off between fidelity and parametric 
complexity / dimensionality. Figure 3 illustrates locations of all control 
points, some coordinates of which are determined directly by the design 
variables, whereas others were either fixed or adjusted automatically to 
maintain surface smoothness. During training, the set of 19 design variable 
values translate to 26 control points that map to some 325 ($=$ 
26$\mathrm{\times }$25/2) design features,\footnote{ The full parametric 
model and mapping of the 19 variables to design features is available from 
the authors.} each representing the distance between a pair of control 
points.
\begin{figure}
    \centering
    \includegraphics[width=1.0\textwidth]{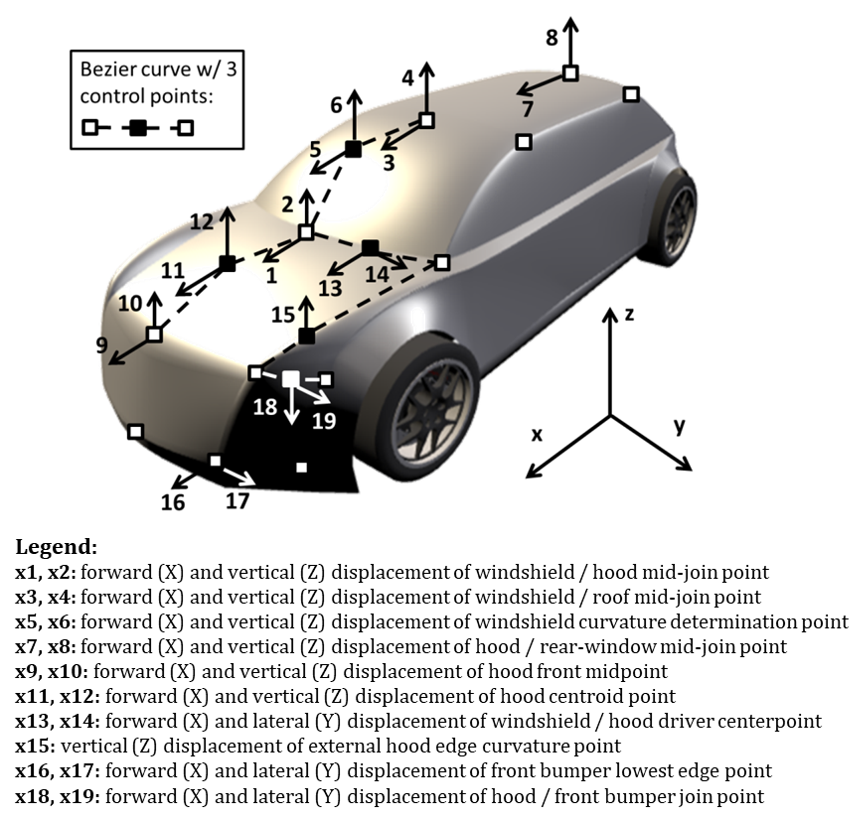}
    \caption{19 Design Variables and Their Control Points}
    \label{fig:fig3}
\end{figure}

There is obviously a cornucopia of functional attributes important to 
potential car buyers. Here, we focus on two of the most critical, price and 
gas mileage (MPG), in terms of their trade-offs to one another and to the 
various form attributes embedded in the design model. In order to avoid 
presumptions of linearity of response, especially to price, we included five 
discrete attribute levels for vehicle price and MPG, selected based on sales 
data of so-called ``CD cars'' (passenger sedans) in the US. Specifically, we 
chose discrete points based on the 10$^{th}$, 25$^{th}$, 50$^{th}$, 
75$^{th}$, and 90$^{th}$ percentiles of both price and MPG, as shown in 
Table 5. Given the coding of both price and MPG into binary level 
indicators, the model as implemented technically encompasses 8 binary 
attributes, as opposed to two used linearly in the ``overall'' preference 
utility function.
\begin{table}[]
    \centering
    \small{
    \begin{tabular}{c c c c}
\hline
Level &	Price (MSRP) &	MPG
(city/highway) &	Percentile
(market data) \\ \hline
1 &	\$23K &	23/27 &	10th \\ \hline
2 &	\$25K &	23/29 &	25th \\ \hline
3 &	\$26K &	24/30 &	50th \\ \hline
4 &	\$29K &	25/31 &	75th \\ \hline
5 &	\$31K &	26/32 &	90th \\ \hline
    \end{tabular}}
    \caption{Function Attribute Levels}
    \label{tab:tab5}
\end{table}

Because the bi-level adaptive technique is novel, it was important to test 
it in theory before doing so in practice. Consequently, we first present an 
extensive series of simulations designed not only to demonstrate parametric 
recovery, but the fit and hit rate performance of each component of the 
bi-level querying technique.

\subsection{Design for Simulation and Empirical Application}
\label{subsubsec:mylabel4}
As proposed at the outset, we presume that the analyst wishes to model both 
form and overall preferences at the individual level, and can pose but a 
limited number of questions via a single-shot survey instrument. Previous 
research (e.g., Sylcott et al. 2013a, Dotson et al. 2019) carrying out 
similar analyses do not lend themselves to such one-shot form vs. function 
preference assessments, due to the need for time-intensive analysis between 
the separate form and function survey instruments. To explore the benefits 
of overcoming this limitation, we simulate three possible modeling options, 
also estimated in the forthcoming empirical application, as shown in Table 
6: Model 1 is the ``base,'' Model 2 is the ``half version'' of the proposed 
model, and Model 3 the ``full version.'' The ``half version'' allows 
assessment of the bi-level question structure and adaptive design 
separately.\footnote{ Because the proposed active learning algorithm was 
built specifically for a bi-level structure and sequential processing, Table 
6 lacks a ``single level {\&} adaptive'' option, so does not have 
2\texttimes 2 factorial structure. To enable comparison, we customized the 
adaptive sampling method to the single-level case, and report these results 
(``Model M1a'') in Table 8 as well. Because they rely on a custom algorithm, 
we do not discuss them further.} 
\begin{table}[]
    \centering
    \small{
    \begin{tabular}{L L L L}
    \hline
\textbf{Models} & \textbf{Querying} &	\textbf{Learning} & \textbf{Sampling} \\ \hline
\textbf{Model 1 (Base: single-level)} &	Single level: 20 purchase questions &	Form and overall:
HB (linear) & Non-adaptive (DOE) \\ \hline
\textbf{Model 2 (Half: bi-level)} &	Bi-level:
10 form questions \& 10 purchase questions &	Form: Rank SVM mix (nonlinear) Overall: HB (linear) & Non-adaptive (DOE) \\ \hline
\textbf{Model 3 (Full: bi-level \& adaptive)} &	Bi-level: 10 form questions \& 10 purchase questions &	Form: Rank SVM mix (nonlinear) 
Overall: HB (linear) & Adaptive \\ \hline

    \end{tabular}}
    \caption{Simulation models and characteristics}
    \label{tab:tab6}
\end{table}
To accord with typical real-world implementations, all three models are 
assumed to be informed by the survey responses of 100 subjects, each with a 
total of 20 questions, including form and purchase questions; an online 
pilot study suggested that 20 questions sufficed for this particular 
application. For validation, we used 100 hold-out questions each for form 
and purchase questions (i.e., 200 total) to compute hit rates. As described 
earlier, ``form'' consisted of 19 continuous design variables, whereas 
``function'' comprised the five levels each for price and MPG.

In line with conventional (non-adaptive) conjoint analysis techniques, Model 
1 (``base'') used purchase questions only, and accommodated both form and 
function within a single linear preference model, which was estimated using 
standard HB methods. For DOE (design of experiments), a Latin hypercube 
sampling method was used to generate questionnaire designs for both 
continuous and discrete variables.\footnote{ Throughout, for Latin hypercube 
sampling, we used the lhsdesign Matlab library; for HB, the hierarchical 
binary logit model in the rhierBinLogit R package (Rossi et al. 2005); and 
for rank SVM, we implemented the rank SVM algorithm based on the LIBSVM 
package (Chang and Lin 2011).}

Model 2 is the ``half version'' of the proposed model, and allows testing of 
the bi-level structure. The bi-level structure makes it possible for form 
and overall preference to be modeled using different specifications -- 
specifically, a nonlinear model for form preference (Gaussian SVM mix model) 
and a linear model for overall preference), after which the form model can 
be nested into the overall model.\footnote{ For the forthcoming online 
experiment (Section 5), subjects may be more easily able to trade-off 
between form and function, because they are shown a pair of vehicle forms 
first, followed by price and MPG with the same forms, an empirical issue not 
easily addressed by simulation alone.} Form preference was estimated by a 
rank SVM mix (Gaussian nonlinear), overall preference using HB (linear), and 
a Latin hypercube sampling method was used for DOE. Compared to Model 1, 
Model 2 sacrifices 10 purchase questions while adding 10 separate form 
questions.

Model 3 is the full version of the proposed model, and tests the bi-level 
structure, non-linear specification, and adaptive questionnaire design 
effects together. The querying and learning structures are the same as Model 
2, but Model 3 uses adaptive sampling so that form and function profiles are 
generated in real time and that each question has different form profiles 
(i.e., a potentially limitless number of forms).

Broadly speaking, the simulation design adapted those used widely in 
academic research applying conjoint methods (e.g., Arora and Huber 2001, 
Toubia et al. 2004, Evgeniou et al. 2005). A mainstay of previous research 
is that response accuracy is controlled by the magnitude of an individual's 
parameters (partworths), while respondent heterogeneity is controlled by the 
variance of parameters (across respondents). We operationalized accuracy and 
respondent heterogeneity by setting each to two levels, ``low'' and 
``high.'' For example, the magnitudes of parameters were set to 
$\mathrm{\beta }=$0.5 and $\mathrm{\beta }=$3 for low and high response 
accuracy, respectively. On a logit scale, these represent deviations in 
log-odds of 0.5 and 3.0 from a baseline of zero (i.e., $\mathrm{\beta 
}=$0); or, in terms of probability, according to $\left( \mathrm{1+}\exp 
\left( \mathrm{-\beta } \right) \right)^{-1}$, which translates into 0.62 
and 0.95, respectively, on a probability baseline of 1/2. The parameter 
variances were set relative to the level of $\mathrm{\beta }$, to $
\mathrm{\sigma }^{\mathrm{2}}\mathrm{=0.5\beta \, }$and $\mathrm{\sigma 
}^{\mathrm{2}}\mathrm{=3\beta \, }$ for low and high respondent 
heterogeneity, respectively. Based on these parameters, four normal 
distributions were defined: $\mathrm{\beta }$ was drawn from each 
distribution, and then four partworth levels for each function attribute, 
($\mathrm{-\beta ,-\beta /3,\beta /3,\beta )}$, were generated, keeping 
constant differences set to 2$\mathrm{\beta }$/3.

For creating individual form preference functions, 19 continuous design 
variables generated a complex form preference model via main-effects 
(``independent'') parameters  $\mathrm{\gamma \, }$and ``interaction'' 
parameters $\mathrm{\delta }$. The independent term of the $k$-th design 
variable, $\mathrm{\gamma }_{k}$  was drawn from four pre-defined 
distributions (analogous to the method used for the function attributes). 
Specifically, for  $k\, =$ 1, 2, ..., 19, four points 
($\mathrm{-}\mathrm{\gamma }_{k}\mathrm{/3}, \mathrm{\gamma 
}_{k}, -\mathrm{\gamma }_{k}, \mathrm{\gamma }_{k}\mathrm{/3)}$  were 
generated, then cubic spline interpolation (denoted $\mathrm{\mathbf{\Phi 
}}\left( \mathrm{\gamma }_{k}\mathrm{,}x_{k} \right))$ was applied to create 
a \textit{continuous} function with respect to the $k$-th design variable, $x_{k}$. We then 
drew 19$\mathrm{\times }$18/2 $=$ 171 interaction terms$,\, \mathrm{\delta 
}_{ij}$, representing the relationship between the $i$-th and $j$-th 
design variables (for $i\ne j)$. The form function, nonlinear but 
continuous, is therefore:
\begin{equation}
\label{eq17}
S\left( \mathrm{\mathbf{x}} \right)\mathrm{\, =\, 
}\sum\nolimits_{k\mathrm{=1}}^{\mathrm{19}} {} \mathrm{\mathbf{\Phi }\, 
}\left( \mathrm{\gamma }_{k}\mathrm{,}x_{k} 
\right)\mathrm{+}\sum\nolimits_{i=1}^{19} \sum\nolimits_{j=1}^{i-1} 
{\mathrm{\delta }_{ij}x_{i}x_{j}} 
\end{equation}
The distributions of $\mathrm{\delta }_{ij}\mathrm{\, }$were balanced in 
the sense that the independent and interaction terms were set to a 2:1 
ratio. [Specifically, following Evgeniou et al. (2005), we randomly 
generated 1000 independent terms and 1000 interaction terms, then compared 
the ratios of absolute values of independent and interaction terms, stopping 
when the standard deviation of the normal distribution for $\mathrm{\delta 
}$ accorded with the 2:1 ratio.] Form score weight, $\mathrm{\lambda }$, in 
Eq. (\ref{eq2}) represents the \textit{importance} of form preference, and was selected to make the 
ratio of absolute values of form score $s$ and function attribute 
preference $\mathrm{\beta }^{T}\mathrm{\mathbf{a}}$ to be 1:2 for the 
``low'' and 2:1 for the ``high'' form importance cases. To do so, we 
generated 10000 random product profiles and 10000 consumer preference 
models, examined the ratio of absolute values of form score and function 
preference, then selected the values that allowed for the 1:2 and 2:1 
ratios.

Consequently, we created eight consumer preference scenarios, as defined in 
Table 7 (note that form score weights are small because form score \textit{values} are 
relatively large). To check hit rate, we generated five sets of all eight 
scenarios, so that 40 total scenarios were used for the simulation.
\begin{table}[]
    \centering
    \small{
    \begin{tabular}{A A A A A A A}
    \hline
    \multirow{2}{2cm}{\centering Form importance} & \multirow{2}{2cm}{\centering Response accuracy} & \multirow{2}{2cm}{ \centering Respondent hetero\-geneity} & 
    \multirow{2}{2cm}{\centering Form score weight ($\lambda$)\footnotemark} & \multicolumn{2}{c}{Form attribute coefficients} & \multirow{2}{2cm}{\centering Functional attribute partworths} \\
    \cline{5-6}
    & & & & \centering Independent terms ($\gamma$) & \centering Interaction terms ($\delta$) & \\ \hline
    \centering Low & \centering Low & \centering 	Low & \centering 	0.0043 &	N(0.5, 0.25) & 	N(0, 4.80) & 	N(0.5, 0.25) \\ \hline
    \centering  Low &\centering 	Low &\centering 	High & \centering 	0.0044 &	N(0.5, 1.5)  &	N(0, 13.7) & 	N(0.5, 1.5) \\ \hline
\centering Low  &\centering 	High  &\centering 	Low  &\centering 	0.0028 &	N(3.0, 1.5) & 	N(0, 56.3) & 	N(3, 1.5) \\ \hline
\centering Low  &\centering 	High  &\centering 	High  &\centering 	0.0057 &	N(3.0, 9.0) & 	N(0, 88.4) & 	N(3, 9.0) \\ \hline
\centering High  &\centering 	Low  &\centering 	Low  &\centering 	0.0173 &	N(0.5, 0.25) & 	N(0, 4.80) & 	N(0.5, 0.25) \\ \hline
\centering High  &\centering 	Low  &\centering 	High  &\centering 	0.0176 &	N(0.5, 1.5) & 	N(0, 13.7) & 	N(0.5, 1.5) \\ \hline
\centering High  &\centering 	High  &\centering 	Low  &\centering 	0.0112 &	N(3.0, 1.5) & 	N(0, 56.3) & 	N(3.0, 1.5) \\ \hline
\centering High  &\centering 	High  &\centering 	High  &\centering 	0.0230 &	N(3.0, 9.0) & 	N(0, 88.4) & 	N(3.0, 9.0) \\ \hline
    \end{tabular}}

    \caption{Consumer Preference Scenarios}
    \label{tab:tab7}
\end{table}
\footnotetext{Form score weights are small owing to scaling. ``True form preference'' was created by combining multiple cubic functions and interaction terms, while ``true function preference'' arises from a simple linear function. Because output values of form preference (form score) are far larger than those for function preference, a small form score weight was used for balance.}

\subsection{Simulation Results}
\label{subsubsec:mylabel5}
Table 8 shows the results of the various simulation scenarios, where hit 
rates were taken as the mean across the five sets. An asterisk (*) indicates 
the best, or not significantly different from best at $p\mathrm{<}0.05$, 
across the three models.
\begin{table}[]
    \centering
    \small{
    \begin{tabular}{p{1cm} p{1cm} p{1.4cm} p{1cm} p{1cm} p{1cm} p{1cm} p{1cm} p{1cm} p{1cm} p{1cm}}
    \hline
    \multicolumn{3}{c}{\textbf{Simulation design}} & \multicolumn{4}{c}{\textbf{Form preference hit rate}} & \multicolumn{4}{c}{\textbf{Overall preference hit rate}} \\ \hline
    \footnotesize{Form importance}	& \footnotesize{Response accuracy}	& \footnotesize{Respondent hetero\-geneity} & Model 1 & Model 1a & Model 2 &	Model 3 &	Model 1 &	Model 1a &	Model 2 &	Model 3 \\ \hline
										
Low	 & Low & 	Low & 	50.8 & 	53.4 & 	65.2 & 	66.2* & 	90.5 & 	90.6 & 	91.9 & 	93.2* \\ \hline
Low & 	Low & 	High & 	51.0 & 	51.1 & 	65.6*	 & 65.3*	 & 91.6 & 	91.7 & 	91.7 & 	90.1 \\ \hline
Low	 & High & 	Low	 & 52.0 & 	53.2 & 	63.3 & 	66.7*	 & 92.7 & 	92.6 & 	93.6*	 & 94.6* \\ \hline
Low	 & High	 & High	 & 51.2 & 	52.0 & 	63.4 & 	65.2*	 & 89.7 & 	90.1 & 	92.3*	 & 92.8* \\ \hline 
High & 	Low & 	Low & 	52.5 & 	52.3 & 	65.1*	 & 66.1*	 & 87.2 & 	87.1 & 	87.9 & 	90.1* \\ \hline
High & 	Low & 	High & 	52.3 & 	52.4 & 	65.2* & 	65.1*	 & 87.2 & 	87.6 & 	88.1*	 & 88.7* \\ \hline 
High & 	High & 	Low & 	53.5 & 	53.6 & 	62.9 & 	66.3*	 & 93.0 & 	93.3 & 	92.8	 & 94.4* \\ \hline 
High & 	High & 	High & 	53.2 & 	53.8 & 	62.4 & 	64.7* & 	87.5 & 	87.6 & 	88.5* & 	89.8*\\ \hline

    \end{tabular}}
    \caption{Simulation Hit Rates}
    \label{tab:my_label}
\end{table}

Except for one case (low form importance, low response accuracy, and high 
respondent heterogeneity), the ``full'' Model 3 outperformed Model 1 for 
both form and overall hit rates. For the form hit rate, every case suggests 
that Model 2 (bi-level structure and non-linear modeling) offers sizable 
improvements over Model 1 (base). Every case also shows Model~3 performing 
as well as or better than Model 2 (adaptive vs. non-adaptive questionnaire 
design), significantly outperforming Model 2 in 5 out of 8 cases. For 
overall hit rate, half the cases favor Model 2 over Model 1. Model~3 
performed as well as or better than Model 2 in all but one case, 
significantly outperforming Model~2 in 3 out of 8 cases. These simulation 
results suggest that the proposed bi-level adaptive method (Model 3) can 
handily outperform the conventional one (Model 1), even with the sacrifice 
of 10 purchase questions. Notably from the perspective of the goals of the 
present study, form preference accuracy can be improved substantially 
(increasing to 65.7{\%} from a base of 52.1{\%}, or a 26{\%} improvement on 
average), enabling marketers to pass along more reasonable target design 
values to industrial designers and engineers.

We conducted several post-analyses to evaluate robustness of results to: 
number of questions, number of attributes, form preference accuracy, 
preference ordering inconsistency, and analyst-tuned parameters, as follows. 
\begin{figure}
    \centering
    \includegraphics[width=\textwidth]{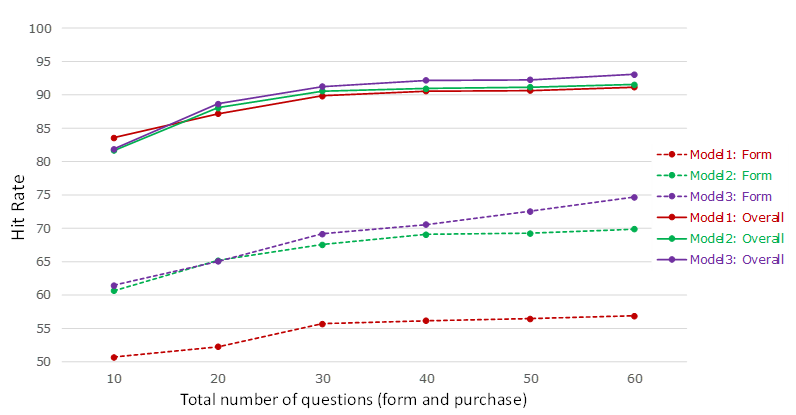}
    \caption{Sensitivity to the Number of Questions}
    \label{fig:fig4}
\end{figure}

{\setstretch{1.12}
\textit{Sensitivity to Number of Questions}. We examine the effects of total number of form and purchase questions, 
from 10 to 60, in increments of 10 on hit rate. Results appear in Figure 4, 
using the results of Model 1 and Model 3 with what is arguably the most 
difficult scenario in Table 8: high form importance, low response accuracy, 
and high heterogeneity. Except for the 10-questions case (i.e., 5 form and 5 
purchase questions for Model 3 vs. 10 purchase questions for Model 1), Model 
3 consistently outperformed Model 1 in overall preference accuracy. This 
owes to the fact that the \textit{form} preference accuracy (for hit rate) of Model 3 was 
always significantly better than for Model 1, \textit{even though half of the purchase questions are sacrificed}. More overall questions 
enabled better form preference accuracy for Model 3, whereas the performance 
of Model 1 did not improve substantially after 30 questions.

\textit{Sensitivity to Number of Functional Attributes}. The second post-analysis examined going from 2 to 10 functional 
attributes, again focusing on the high form importance, low response 
accuracy, high heterogeneity scenario. Although including 10 attributes is 
sometimes feasible (though highly taxing for respondents) in a real-world 
online study, the proposed query generation procedure would entail a 
higher-dimensional discrete optimization problem, potentially hindering 
real-time performance. We focus on Models 1 and 3 for a comparison across 
bi-level structure: results suggest going from 2 to 10 attributes has a 
small effect on form preference hit rate for Model 1 (52.3 to 51.5), but 
none for Model 3 (65.0 to 65.0). However, overall hit rate degrades sharply: 
Model 1, 87.2 to 51.5; Model 3, 88.7 to 54.6. Although this is necessarily 
speculative, form preference accuracy in Model 3 appears unaffected by 
number of functional attributes, owing to its form preference being trained 
separately using bi-level structure. 

\textit{Form Preference Accuracy}. Ideally, one would also like to assess parametric recovery for the true 
vs. estimated form preference models. But this is not directly possible, 
because the true and estimated functions have different types and numbers of 
parameters: the former created by combining cubic curves and interaction 
terms, the latter an estimated SVM whose parameters are Lagrangian 
multipliers. However, it is possible to compare using hit rate as a 
``scale-free'' measure of prediction performance. Specifically, we compare 
form importance between the true and estimated models across individuals by 
using RMSE, with form importance calculated as the ratio the \textit{range} of form 
preference (value after multiplying by form score weight, $\lambda 
_{i}$s$_{i}$, in Eq. (\ref{eq1}) to range of function preference. For brevity, we 
do this for the same ``difficult'' scenario in Table 8 as before (high form 
importance, low response accuracy, and high heterogeneity). RMSE values are 
2.141 for Model 1, 0.149 for Model 2, and 0.119 for Model 3, with all 
pairwise differences significant at $p$ \textless .005. As such Model 3 is very 
strongly favored over the other two, with very inferior form prediction 
performance for Model 2. 

\textit{Robustness to Respondent ``Noise'' in Preference Ordering}. Because the hard-margin SVM lacks a dedicated error structure, one might 
presume it could ``break down'' in the case of contradictory preference 
orderings, for example, a consumer who effectively asserts A \textgreater B 
and, elsewhere, B \textgreater A. To assess this possibility, we conducted 
additional simulations adding a `noise factor', i.e., probability that the 
customer reverses preference orderings. Specifically, for each form 
question, contradictory preference ordering occurs randomly, with 
probability 10{\%} or 20{\%}, and we compared both form and overall 
preference hit rate with the base (0{\%}, i.e., non-contradictory) case, as 
follows. Overall, there is a moderate fall-off in form preference hit rate, 
but very little in overall preference. For Model 3, form preference hit rate 
is \textbraceleft 65.1, 62.5, 58.6\textbraceright for \textbraceleft 0{\%}, 
10{\%}, 20{\%}\textbraceright contradictory preference order probabilities, 
while overall preference -- which includes the effect of covariates -- is 
affected far less, \textbraceleft 88.7, 88.0, 87.8\textbraceright . In 
short, the hard-margin SVM provided reasonable performance in the face of 
even moderate preference inconsistency.

\textit{Robustness to }$c_{j}$. Lastly, as mentioned earlier, we examine robustness to the choice 
of ``gaps'' in (\ref{eq3}), that is, $c_{j} $being set to 1 (better) and 2 (much 
better). In Appendix G, we explain how this sensitivity can be assessed 
using the Lagrange multipliers of the solution of the dual problem in 
(\ref{eq4}), and carry out two simulations: increasing the values of $c_{j}$ while 
holding them in fixed 1:2 ratio (\textbraceleft 0.1,0.2\textbraceright to 
\textbraceleft 1000,2000\textbraceright ), and altering the ratio itself 
(1:2 to 1:10). Results suggest almost complete insensitivity of form 
preference hit rate across all these scenarios.

}

\subsection{Bi-Level Query Performance}
The bi-level query method offers two advantages. First, from a learning 
(parameter estimation) perspective, note that while form preference is a 
nonlinear function on a high-dimensional parameter space, overall utility 
can be expressed as a linear-in-parameters function of form preference and 
functional attributes (as well as interactions, although we tested for and 
did not find these for our application). By using SVM for form preference 
and HB for choice, the bi-level method exploits this structure, and should 
have better generalization performance than a single-level model (and this 
is consistent with our forthcoming application, e.g., Table 10, Model 1 vs. 
2). Secondly, from a query perspective, the form-only queries are used to 
estimate form preference, which supports the choice of informative purchase 
query, and this salutary effect on query selection is reflected in the 
comparison between Models 2 and 3. And to reiterate, as shown in Appendix F, 
the performance of sequential query generation, vs. simultaneous, is 
superior for a fixed budget.

However, simulation on synthetic data and an empirical application, no 
matter how compelling, do not constitute a theoretical guarantee. To this 
end, in Appendix A, we present a detailed formal proof addressing when and 
why the bi-level query is superior. It relies on the computation of 
generalization error for the single- and bi-level cases, and has two 
overarching conclusions: (1) When the form signal-to-noise ratio 
$\mathrm{\vert \vert }\beta \mathrm{\vert }\mathrm{\vert 
}_{\mathrm{1}}\mathrm{/}\sigma_{s}$ is sufficiently large, and the 
form-to-utility noise ratio $\sigma_{s}^{\mathrm{2}}\mathrm{/}\sigma 
_{y}^{\mathrm{2}}$ is sufficiently small, the bi-level questionnaire is 
superior; (2) In the extreme case where form responses are noiseless 
($\sigma_{s}^{\mathrm{2}}\mathrm{=0})$, the bi-level query is always 
better. 

\section{Online Experiment}
\label{subsec:mylabel8}
The simulation spoke clearly to the advantages of bi-level adaptive 
querying. But, as the saying goes, what works well in theory may not do so 
in practice. To assess real-world performance of the bi-level adaptive 
technique for form preference optimization, we conducted three online 
surveys that correspond with the models simulated in Table 8. Three online 
groups were recruited through ClearVoice Research, a prominent online panel 
provider, and, to accord with the simulation scenarios, each comprised 100 
subjects. Demographics were specified to match with the general US adult 
population of car-owning households; post-analysis confirmed the accuracy of 
recruitment.\footnote{ Averages were as follows: 50.2 years age; 82{\%} 
Caucasian; 81{\%} suburban or small town; 95{\%} high school; 69{\%} some 
college; 58{\%} working; 4{\%} student; 15{\%} homemaker; 23{\%} retired; 
{\$}58767 household income; 55{\%} married; 4.3 family size; 2.6 children; 
65{\%} spouses employed. Full cross-classified categorical breakdowns are 
available from the authors.} A total of 20 questions were used for learning 
and 10 holdout questions (i.e., 5 form and 5 purchase holdout questions) 
used to check hit rates. Respondents completed the online task of their own 
volition, with no time limits, on devices of their choosing. The survey 
mechanism was implemented as follows. On the client side, JavaScript, WebGL 
and ThreeJS were used to enable real-time 3D model rendering and interaction 
through mainstream web browsers, with no additional software requirements. 
Critically, users were able to rotate the real-time-generated 3D images for 
each presented form before deciding on their responses. On the server side, 
Google App Engine was used for both executing real-time machine learning 
algorithms and for data storage. [See Appendix H for details on the query 
engine platform and resultant respondent response time tests.]

\subsection{Parameters and Model Performance}
All three models were estimated as described in Sections 2 and 3. Parameter 
estimates for form and function attributes appear in Table 9, and we discuss 
implications about their natural groupings after comparing relative 
performance quality.\footnote{ As mentioned earlier, interactions can be 
included in the overall utility model. We performed an exhaustive search 
over the homogeneous model space -- where our model is nested in those with 
interactions -- and a targeted search, via Stan (https://mc-stan.org/), over 
the heterogeneous one. We saw no ``significant'' interactions, inputting 
Price and MPG either as binary levels (4 df each) or as values (1 df each). 
Specifically, for the latter, 95{\%} HDRs are, with all variables 
mean-centered: Styling*Price: [-0.039, 0.502]; Styling*MPG: [-0.665, 0.370]; 
Price*MPG: [-0.028, 0.098]. Full interaction model details are available 
from the authors.} 
\begin{table}[]
    \centering
    \small{
    \begin{tabular}{p{1.4cm} | p{0.9cm} | p{0.9cm} p{0.9cm} p{0.9cm} p{0.9cm} p{0.9cm} | p{0.9cm} p{0.9cm} p{0.9cm} p{0.9cm} p{0.9cm}}
    \hline
 	& \textbf{Form}	& \multicolumn{5}{c}{\textbf{Price}} & \multicolumn{5}{c}{\textbf{MPG (city/highway)}} \\
 	\cline{2-12}
 	& $\lambda_i$ &	\$23K &	\$25K &	\$26K &	\$29K	& \$31K	& 23/27	& 23/29	 & 24/30 &	25/31 & 26/32 \\
 	\cline{2-12}
Mean &		4.45 &		1.06 &		0.42 &		0.21 &		-0.83 &		-0.86 &		-0.95 &		-0.50 &		0.13 &		0.64 &		0.68 \\
StdErr  &		0.25 &		0.14 &		0.08 &		0.07 &		0.11 &		0.11 &		0.10 &		0.08 &		0.07 &		0.07 &		0.08 \\
Heterog. &		1.66 &		0.89 &		0.46 &		0.36 &		0.76 &		0.77 &		0.53 &		0.37 &		0.26 &		0.40 &		0.41 \\
$r$ with $\lambda_i$  &		--- &		-0.22 &		-0.39 &		-0.45 &		0.35 &		0.35 &		0.40 &		0.34 &		0.10 &		-0.46 &		-0.44 \\
\hline

    \end{tabular}}
    \caption{Parameter Estimates and Summaries}
    \label{tab:tab9}
\end{table}

Table 9 lists parameter means, standard errors, estimated random 
coefficients standard deviation (i.e., of the heterogeneity distribution), 
and random effects correlation ($r)$ with the Form score. The Form score is 
clearly ``significant'', on average, and examination of the posteriors for 
individual $\lambda_{i}$ suggests form is significantly (.05) positive for 
more than \textthreequarters of the participants. In other words, ``Form 
Matters'' not just as an overall parameter mean, but for most people 
individually. The last row suggests that -- generally speaking -- people -- 
who value Form also ``prefer'' higher prices and lower MPG, both of which 
are consistent with a higher WTP overall, and perhaps a larger automotive 
budget. [Note that, because of the zero-mean scaling for the Price and MPG 
partworths within-person, interpreting their significance levels and 
heterogeneity distributions is less straightforward than for Form.] 

Hit rates for the three models in the online experiments are shown in Table 
10, with the proposed model performing best across the board.
\begin{table}[]
    \centering
    \begin{tabular}{r c c}
    \hline
 	& Form preference hit rate & Overall preference hit rate\\ 
 	\hline
Model 1: Single-Level (Base) & 54.4\% &	57.2\% \\
\hline
Model 2: Bi-Level (Half) & 62.0\%* & 62.0\% \\
vs. Model 1	& (7.6\%)	& (4.8\%) \\
$p$ against M1 &	0.015 &	0.111 \\ \hline
Model 3: Bi-Level Adaptive (Full) &	64.0\%*	& 68.2\%* \\ 
vs. Model 1	& (9.6\%) &	(11.0\%) \\
$p$ against M1 &	0.003 &	0.001 \\
$p$ against M2 & 	0.537 &	0.027 \\ \hline
\multicolumn{3}{c}{\footnotesize{Figures in parentheses show percentage improvement over ``base'' Model 1}} \\
\multicolumn{3}{c}{\footnotesize{
*Best, or not significantly different from best, at p<0.05, across all models}
}

    \end{tabular}
    \caption{Hit Rates in Online Experiment}
    \label{tab:tab10}
\end{table}

Model 2 (in Table 10) clarifies the effect of the bi-level structure, which 
entails substantial improvements in prediction, an increase of 7.6{\%} and 
4.8{\%}, for form and overall preferences hit rates, respectively, compared 
to Model 1 (or 14.0{\%} and 8.4{\%} of their respective baselines). Model 3 
(again in Table 10) shows the effect of the bi-level structure as before, 
but also of adaptive sampling. These results further suggest that Model 3 
offers an increase of 9.6{\%} and 11.0{\%} for form and overall preferences 
hit rates, respectively, compared to Model 1 (or 17.6{\%} and 19.2{\%} of 
their respective baselines) and an increase of 2.0{\%} and 6.2{\%} compared 
to Model 2 (or 3.2{\%} and 10.0{\%} of baseline). The overall pattern of 
results suggests that adaptive sampling is useful to elicit both non-linear 
form preferences and linear overall preferences. Specifically, the bi-level 
structure appears to have affected predictive accuracy for form preference 
more than for overall preference; and adaptive sampling affected overall 
preference predictions more than those for form.

Although the overarching purpose of this study is to model both form and 
function preferences \textit{together}, within the confines of a one-shot survey, and to 
measure the trade-offs among specific design variables and functional ones, 
we did test another model that did not incorporate form. Specifically, we 
removed form attributes from Model 1 to check overall preference prediction 
based on functional attributes alone. In ``Model 1a'', we trained the 
overall preference model using only the function attributes, price and MPG, 
then re-checked hit rate. The results were dramatic: the hit rate increases 
to 64.6{\%}, from the 57.2{\%} of Model 1 (or 12.9{\%} of baseline). This 
suggests that \textit{predicting overall preference by incorporating form design variables and function attributes within a single linear model may be suboptimal as a general approach}. Model 2 in fact shows slightly poorer performance in overall 
preference hit rate, as it sacrifices 10 purchase questions and instead 
models form preference. The proposed method, Model 3, by contrast, affords 
significantly better prediction (68.2{\%}) for overall preference than Model 
1a (64.6{\%}).

\subsection{Segmentation via Form and Function Part-Worths}
An examination of parameter estimates and summaries in Table 9 suggests that 
Form is important, and nontrivially correlated with ``function'' attribute 
levels; but these are coarse metrics of their interrelation, and don't hint 
at what to actually produce for a heterogeneous market. In this vein, we 
first examine whether there appear natural groupings -- that is, a 
segmentation -- within the data in terms of the overall ``weight'' placed on 
form and on the two functional attributes (price and MPG). Given their 
within-respondent zero-sum scaling, Price and MPG importances can be 
calculated by the difference between the highest and lowest partworth; these 
values and \textbraceleft $\mathrm{\lambda }_{i}$\textbraceright are 
averaged across MCMC draws to compute three deterministic values for each of 
the 100 subjects, who can then be clustered using (first) hierarchical and 
(subsequently) K-means methods according to their form ($\mathrm{\lambda 
})$, price, and MPG importances. Typical metrics suggested four clusters fit 
the data best; both raw and standardized averages appear in Table 11.
\begin{table}[]
    \centering
    \begin{tabular}{c c c c c c}
    \hline
	& \textbf{Overall} &	\textbf{Group 1} &	\textbf{Group 2} &	\textbf{Group 3} &	\textbf{Group 4} \\
	\cline{2-6}
\textbf{Size} &	\textbf{100\%}	& \textbf{24\%} &	\textbf{23\%} &	\textbf{14\%} &	\textbf{39\%} \\
    \cline{2-6}
	\multicolumn{6}{c}{Raw Differences} \\ 
	\cline{2-6}
\textbf{Form} &	\textbf{4.45} &	2.92 &	5.42 &	2.26 &	5.62 \\
\textbf{Price} &	\textbf{1.92} &	4.09 &	2.69 & 	0.71 &	0.55 \\
\textbf{MPG} &	\textbf{1.63} &	2.24 &	2.00 &	2.00 &	0.92 \\
    \cline{2-6}
	\multicolumn{6}{c}{Standardized Differences} \\
	\cline{2-6}
\textbf{Form} &	\textbf{0} &	-0.92 &	0.58 &	-1.31 &	0.70 \\
\textbf{Price} &	\textbf{0} &	1.33 &	0.47 &	-0.73 &	-0.83 \\
\textbf{MPG} &	\textbf{0} &	0.67 &	0.40 &	0.40 &	-0.79 \\
\hline

    \end{tabular}
    \caption{Clustering Based on Form, Price, MPG}
    \label{tab:tab11}
\end{table}

The four clusters can thereby be roughly interpreted as:

\indent \indent \textbf{Group 1: Price} and \textbf{MPG} are important (relative to Form) \\
\indent \indent \textbf{Group 2: All three }(Price, MPG, Form) are valued in balance \\
\indent \indent \textbf{Group 3: MPG} is important (relative to Price especially) \\
\indent \indent \textbf{Group 4: Form} is \textbf{very} important (relative to Price and 
MPG)

Of the four groups, the fourth is far more concerned with vehicle aesthetics 
than the other three, while the first group doesn't appear to value Form 
very strongly. In other words, \textit{willingness to pay for vehicle form} is high in group 4 and low in group 1. Group 
3, however, has relatively high WTP, since both Form and MPG are valued 
\textit{relative} to Price.
\begin{figure}
    \centering
    \includegraphics[width=\textwidth]{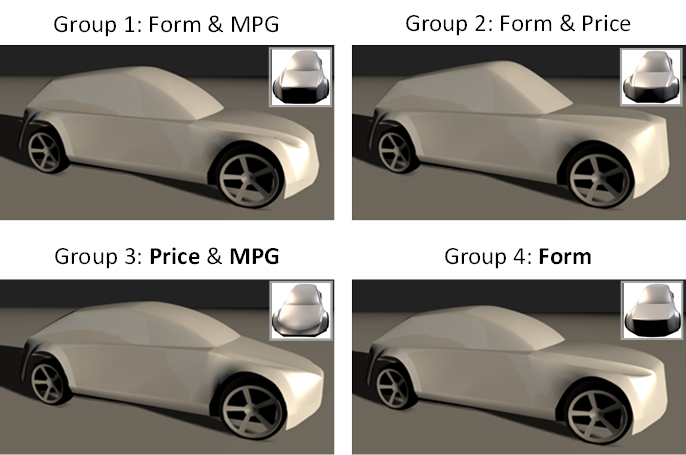}
    \caption{Optimal Designs for Four Extracted Clusters}
    \label{fig:fig5}
\end{figure}
But we can, of course, do more: because the underlying model is literally 
built around the idea of real-time visual generation, it can render the car 
designs ``most liked'' by each of the four groups. These appear, from the 
sideways and head-on perspective, in Figure 5, i.e., the ``forms'' along 
with the ``functions'' most valued by each group. Although this amounts to 
visual inspection, note that the design for Group 3 resembles the Toyota 
Prius: it has relatively streamlined silhouette, e.g., the transition from 
the hood to the front windshield. This is consistent with Group 3's 
weighting MPG the most, especially in comparison with Price. Similarly, the 
design for Group 4 has lower profile than those of Groups 1 and 2, 
consistent with their finding form very important. While the designs 
themselves may not look radically different, this is to be expected, for 
several reasons. First, our parametric model does not incorporate every 
element of car design (e.g., door and window shape, metal vs. plastic 
exterior panels) nor common attributes (e.g., color, audio, interior 
configuration). Second, the extracted designs indeed span a rather large 
swath of the product topology space for \textit{this particular type of car}, as can be gleaned from examining 
the range of design variables and control points (to which we return later). 
And finally, we must keep in mind that these four groups, while clustered 
according to preference, still do not have \textit{identical} \textit{within-group preferences}. That is, the rendered designs 
are for the ``centroid'' of each group; it is possible using the model to 
design for smaller groups, or even individuals, if the analyst so wishes.

Regardless, once the model is run and both form and function parameters are 
obtained, it is possible to extract much more than purportedly homogeneous 
segments that parcel consumers on their valuation of design \textit{overall} (vs. price and 
MPG). Rather, the analyst can use individual-level price coefficients and 
weights on the underlying control points to infer which \textit{specific design elements} individuals, or 
groups, are willing to pay for. Product designers can then compare the cost 
of provision of those design elements -- say, a more sloped profile that 
would provide less space for the engine compartment but could enhance 
aerodynamics -- with WTP for that element, as well as any available 
demographics that could serve as a classic discriminant function or 
hierarchical model covariate. We conclude our discussion of the empirical 
conjoint data with an analysis of this sort: which design elements are 
associated with relatively high WTP, and do these vary substantially across 
the respondent pool?

\subsection{Trade-offs between Specific Form and Function Attributes}
To examine the relationship between functional attributes like price and 
\textit{specific} form attributes is somewhat more complex than the usual trade-off 
computations that typically follow conjoint. This is because, while most 
functional attributes in conjoint are ``vector'' type -- e.g., all else 
equal, it is better to get \textit{higher} mileage and haggle for a \textit{lower} price -- this is 
seldom the case with design attributes. For example, one might like a large 
and highly angled windshield, but both size and pitch are self-limiting: no 
matter how much one might like more of them, each eventually veers into 
dysfunctionality. That is, design attributes tend to be of the U-shaped 
ideal-point type, with a respondent-specific internal ``Goldilocks'' 
maximum. Because we have calibrated an individual-preference model, it is 
not difficult to calculate, for each respondent, four quantities: a maximum 
(what degree of that element is liked best, contingent on all the other form 
elements being jointly optimized), its form ``score'' (as per Eq. (\ref{eq1}) {\&} (\ref{eq2})), 
and the associated gradient and Hessian. The latter two can be quickly 
computed using numerical techniques, and for each respondent; and stable 
quantities were obtained for all four design variables. 

\textit{Sensitivity to Design Variables}. The diagonal components of the Hessian -- which will all be negative for 
internal maxima -- correspond to curvature or sensitivity: how (un)willing 
the respondent would be to give up one unit of that form attribute? [Recall 
that all form elements were standardized before optimization, rendering them 
comparable on a dimensionless scale.] 
\begin{table}[]
    \centering
    \scriptsize{
    \begin{tabular}{|c !{\vline width 1pt} r !{\vline width 1pt} B B B B B | B B B B B|}
    \hline
    \multirow{2}{1cm}{\centering \textbf{Design Vars.}} & 
    \multirow{2}{1.1cm}{\centering \textbf{Hessian at Max}} &
    \multicolumn{5}{c|}{\textbf{WTT: Price (\$1000; Median)}} &
    \multicolumn{5}{c|}{\textbf{WTT: Mileage (1 MPG; Median)}} \\
    & & \textbf{Overall}	& \textbf{Group 1} &	\textbf{Group 2}	& \textbf{Group 3} &	\textbf{Group 4} &	\textbf{Overall}	 & \textbf{Group 1}	& \textbf{Group 2} &	\textbf{Group 3}	& \textbf{Group 4} \\
    \hline
\textbf{x1}	& 	-33.20	& 	\textbf{5.36}	& 	0.67	& 	2.40	& 	4.87	& 	11.71	& 	\textbf{1.52}	& 	0.64	& 	1.36	& 	0.69	& 	3.76\\
\textbf{x2}	& 	-36.90	& 	\textbf{4.86}	& 	0.62	& 	2.89	& 	6.44	& 	13.47	& 	\textbf{1.76}	& 	0.64	& 	1.53	& 	0.98	& 	3.40\\
\textbf{x3}	& 	-12.40	& 	\textbf{2.10}	& 	0.20	& 	1.24	& 	2.93	& 	5.49	& 	\textbf{0.67}	& 	0.20	& 	0.48	& 	0.40	& 	1.70\\
\textbf{x4}	& 	-6.10	& 	\textbf{0.86}	& 	0.07	& 	0.35	& 	1.21	& 	2.71	& 	\textbf{0.31}	& 	0.08	& 	0.17	& 	0.24	& 	0.86\\
\textbf{x5}	& 	-2.50	& 	\textbf{0.36}	& 	0.04	& 	0.21	& 	0.55	& 	1.41	& 	\textbf{0.12}	& 	0.04	& 	0.10	& 	0.08	& 	0.39\\
\textbf{x6}	& 	-3.60	& 	\textbf{0.67}	& 	0.06	& 	0.32	& 	0.94	& 	1.85	& 	\textbf{0.20}	& 	0.06	& 	0.14	& 	0.14	& 	0.51\\
\textbf{x7}	& 	-6.40	& 	\textbf{0.80}	& 	0.12	& 	0.42	& 	0.58	& 	2.64	& 	\textbf{0.25}	& 	0.10	& 	0.23	& 	0.18	& 	0.81\\
\textbf{x8}	& 	-0.20	& 	\textbf{0.02}	& 	0.00	& 	0.01	& 	0.14	& 	0.08	& 	0.01	& 	0.00	& 	0.01	& 	0.02	& 	0.02\\
\textbf{x9}	& 	-36.30	& 	\textbf{4.20}	& 	0.60	& 	2.65	& 	3.55	& 	13.46	& 	\textbf{1.65}	& 	0.58	& 	1.59	& 	0.63	& 	4.12\\
\textbf{x10}		& -38.30	& 	\textbf{5.30}	& 	0.87	& 	2.79	& 	5.95	& 	13.71	& 	1.81	& 	0.57	& 	1.72	& 	0.70	& 	4.65\\
\textbf{x11}	& 	-1.90	& \textbf{	0.17}	& 	0.03	& 	0.16	& 	0.44	& 	1.16	& 	\textbf{0.08}	& 	0.02	& 	0.08	& 	0.05	& 	0.36\\
\textbf{x12}	& 	-11.30	& 	\textbf{1.41}	& 	0.25	& 	0.87	& 	1.93	& 	4.38	& 	\textbf{0.56}	& 	0.13	& 	0.57	& 	0.23	& 	1.21\\
\textbf{x13}	& 	-18.20	& 	\textbf{1.94}	& 	0.38	& 	1.25	& 	1.91	& 	7.21	& 	0.89	& 	0.36	& 	0.74	& 	0.30	& 	2.02\\
\textbf{x14}		& -34.40	& 	\textbf{4.92}	& 	0.69	& 	2.37	& 	4.72	& 	16.46	& 	\textbf{1.76}	& 	0.70	& 	1.48	& 	0.65	& 	3.60\\
\textbf{x15}	& 	-10.30	& 	\textbf{0.84}	& 	0.27	& 	0.76	& 	1.23	& 	5.14	& 	\textbf{0.41}	& 	0.19	& 	0.39	& 	0.12	& 	1.10\\
\textbf{x16}	& 	-17.70	& 	\textbf{2.25}	& 	0.48	& 	0.81	& 	2.36	& 	6.21	& 	\textbf{0.78}	& 	0.40	& 	0.58	& 	0.36	& 	2.17\\
\textbf{x17}	& 	-6.20	& 	\textbf{0.80}	& 	0.21	& 	0.55	& 	0.73	& 	3.34	& 	\textbf{0.29}	& 	0.15	& 	0.26	& 	0.09	& 	0.85\\
\textbf{x18}	& 	-17.80	& 	\textbf{2.64}	& 	0.38	& 	1.29	& 	2.19	& 	8.27	& 	\textbf{1.12}	& 	0.40	& 	0.76	& 	0.31	& 	2.08\\
\textbf{x19}		& -32.80	& 	\textbf{4.45}	& 	0.70	& 	2.36	& 	4.76	& 	14.95	& 	\textbf{1.46}	& 	0.53	& 	1.18	& 	0.64	& 	4.36	\\
\hline
\textbf{Average}	& \textbf{-17.20}	& 	\textbf{2.31}	& 	\textbf{0.35}	& 	\textbf{1.25}	& 	\textbf{2.50}	& 	\textbf{7.04}	& 	\textbf{0.82}	& 	\textbf{0.31}	& 	\textbf{0.70}	& 	\textbf{0.36}	& 	\textbf{2.00}\\
\hline
\textbf{MPG}	& 	---		& \textbf{2.41}	& \textbf{0.97}	& 	\textbf{1.73}	& 	\textbf{6.29}	& 	\textbf{3.11}	& 	1	& 	1	& 	1	& 	1	& 	1\\
\hline

\multicolumn{12}{p{15.5cm}}{}\\
\multicolumn{12}{p{15.5cm}}{\centering \scriptsize{Reported values for Willing To Trade-off Price (WTTP) and Mileage (WTTM) refer to units of \$1000 or miles, respectively, for a unit change in the design variable (calculated as a .01 change, multiplied by 100). These thus provide a ``local'' approximation of the full range of the design variables, each of which was pre-standardized. Design variable descriptions appear in Figure 3. Specifically, the final line’s median WTP in dollars for each additional MPG is \$2410 (overall), and \$970, \$1,730, \$6,290, \$3,110 for Groups 1-4, respectively.}}\\

    \end{tabular}}
    \caption{Form vs. Function Trade-offs}
    \label{tab:tab12}
\end{table}

For each of the original 19 design variables (see Figure 3), summary 
statistics for these sensitivities appear in Table 12, where it is apparent 
that there is a wide variance across both design attributes and people. 
\textit{Generally speaking, ``front and center'' design variables were valued far more than those harder to see while driving}. For example, x8 -- elevation of the central point where the back 
windshield meets the roof -- appears to be the least sensitive design 
element: its median value was -0.169 (tabled value $=$ -0.2), compared to 
the average (across all design variable medians) of -17.2. In simple terms, 
respondents were, on average, 100 times less sensitive to this design 
parameter than the others. By contrast, six design attributes stand out in 
terms of high sensitivity to change, listed with their medians (see Figure 
3): x1 (-33.2) and x2 (-36.9), the horizontal and vertical position of the 
midpoint the hood/windshield join; x9 (-36.3) and x10 (-38.3), the 
horizontal and vertical position of the midpoint of the hood front; x14 
(-34.4), the lateral displacement of the hood/windshield join point directly 
in front of the driver; and x19 (-32.8), the outward displacement of the 
driver's side hood front. 

\textit{Trade-offs Against Design Variables}. However, examining sensitivities alone is merely suggestive: perhaps the 
consumers who are most sensitive to \textit{specific} elements of design also have the lowest 
\textit{overall} value for design, relative to price (or MPG). It is even possible that 
nearly all respondents, while responding to design changes \textit{in and of themselves}, have no 
trade-off against ``functional'' attributes, especially price. As such, we 
wish to construct metrics for ``Willingness To Trade-off'' (WTT) for each 
design attribute vs. the two functional attributes (price and mileage) in 
our study, as well as the functional attributes against one another, a 
standard ``WTP'' calculation in traditional conjoint (Sonnier et al. 2007). 
This is made more complex by our having allowed for nonlinear response to 
the five levels of both price and MPG. So, for simplicity of presentation, 
we compute two values, akin to the interquartile range, for each consumer's 
utility function: the difference in the 25$^{th}$ and 75$^{th}$ percentile 
values. That is, the partworth differences for {\$}29K vs. {\$}25K and 23/29 
MPG vs. 25/31; or, more simply {\$}4000 and 2MPG, which can then be 
standardized into willingness-to-trade-off {\$}1000 and 1MPG (by dividing by 
4 and 2, respectively). Finally, the deterministic part of the utility model 
is given as $\mathrm{\lambda }_{i}s_{i}\mathrm{\, +\, }\mathrm{\beta 
}_{i}^{T}\mathrm{\mathbf{a}}$, so that we can unambiguously answer ``what 
{\%} change in each design variable within $s_{i}$ maintains utility if 
price were either {\$}1000 higher (WTT Price, WTTP) or mileage 1 MPG better 
(WTT Mileage, WTTM)?'' For each consumer, this entailed both the ``form'' 
model for $s_{i}$ and the estimated value of $\mathrm{\lambda }_{i}$, 
which measures overall design importance. 

As a check, we first computed the WTP for MPG, a standard trade-off between 
conjoint attributes for automotive applications. The literature reports a 
wide range of values, depending on which sort of cars are included, the 
method of task (e.g., conjoint, purchase data, or field experiment), 
demographic composition of the respondent pool, and, critically, the range 
of prices and MPG values studied. Greene's (2010) review of this literature 
highlights findings from Gramlich (2010), who found that WTP in {\$}/mile 
(calculated based on an increase from 25 to 30 MPG and a gas price of 
{\$}2/gal.), was approximately {\$}800 for luxury cars, {\$}1480 for compact 
cars, and {\$}2300 for SUVs; but that this rose {\$}1430, {\$}2580, and 
{\$}4100, respectively, for a gas price of {\$}3.50/gal (in 2008 prices). In 
our study (last row of Table 12), for mid-priced sedans, median WTP in 
{\$}/mile was {\$}2410,\footnote{ Specifically, in this case, $\Delta 
$(partworths of {\$}25K and {\$}29K)/$\Delta $(partworths of 25/31MPG and 
23/29MPG), rendered in {\$}1/mile. Sonnier, Ainslie {\&} Otter (2007) 
provide additional detail on such calculations.} well within the latter 
reported range, lending external validity to our results.

Our goal was to quantify trade-offs between our 19 design variables and both 
price and MPG. To do so, we computed median values for willingness to trade 
off a .01 deviation (from optimum) in each design variable against price 
(WTTP, in {\$}1000) and mileage (WTTM, in 1~MPG); these were multiplied by 
100 to reflect the relative size of the .01 deviation to the normalized unit 
scaling for each of the 19 design variables. Results appear in the last four 
columns of Table 12. Because the trade-off between price and MPG was 
{\$}2410, we would expect the values for MPG to be about 40{\%} of those for 
price (based on the last row of Table 12 the exact value is 35.6{\%}). 

It is apparent that some design variables were far more valued than others, 
roughly tracking with the results for the Hessian (column 2 of Table 12). 
For example, as before, x1 and x2 (horizontal and vertical position of the 
hood/windshield join midpoint) were each valued by approximately {\$}50 for 
each .01 deviation from optimum (tabled values, 5.36 and 4.86, respectively, 
in {\$}1000 units/100), and would correspond with mileage losses of .0152 
and .0176 MPG each. These two (x1, x2) were not alone, as several of the 
design variables showed similar substantial sensitivity, with .01 changes 
corresponding to approximately {\$}50 in price (e.g., x9, x10, x14, x19). 

Using Table 12, one can tally up medians (i.e., summing columns 4 and 9, 
then dividing by 100) to calculate an ``omnibus'' value for design changes 
\textit{overall}, based on .01 deviations from each consumer's optimum design along all 19 
dimensions. Doing so translates into approximately {\$}439 in WTTP and .157 
miles in WTTM, both substantial values, given the small 1{\%} deviations. Of 
course, the underlying choice model, based on form utility, is highly 
nonlinear, so one must take care in extrapolating such ``local'' results to 
the entire design space, where whole regions are likely to show little slope 
due to their being non-viable for particular consumers (e.g.,~designs they 
actively dislike). Furthermore, even with substantial heterogeneity, the 
availability of individual-level results means that -- given the cost of 
production of different design elements and demographics for respondents -- 
a manufacturer could roughly compute whether a certain \textit{kind }of consumer would be 
willing to pay the added cost of a proposed design alteration, or whether it 
might be worth reduced fuel efficiency or trade-off against any traditional 
(functional) conjoint attribute.

\subsection{Form and Function Trade-Offs for Group-Optimized Designs}
One advantage of the model, as detailed in Section 5.2 and illustrated in 
Figure 5, is the ability to form groupings based on form and function 
preference, and determine ``high utility'' -- that is, visually appealing -- 
designs for each group. But model output also allows specific form vs. 
function trade-offs to be assessed for each design; these are also broken 
out, by median, in Table 12. These not only help characterize each group's 
core trade-offs, but indicate whether there is substantial heterogeneity in 
terms of valuation of each design element. Such an assessment is critical, 
for example, in the literature on component commonality in flexible 
manufacturing (e.g., Fixson 2007): groups with low interest in, or WTP for, 
a particular form or functional element can often be provided with a 
relatively high-quality version by adopting or adapting existing designs and 
benefiting from production economies-of-scale.

Table 12 lists WTT Price and Mileage for the various design elements by 
Group, as well as the trade-off between MGP and Price. Results indicate 
substantial heterogeneity, with Group 4 -- for whom Form is very important 
overall -- having highest WTP on average: roughly triple the entire-sample 
value (7.04 vs. 2.31), although they don't have highest WTP for every 
element. By contrast, Group 1 -- who value Form little compared with Price 
and MPG -- have low WTP for nearly all design elements, yet are quite 
sensitive, with WTPs of {\$}600 and {\$}870, respectively, across the range 
of design variables x9 and x10, which concern the forward and vertical 
displacement of hood front midpoint. While considering every such form 
trade-off for the four groups would take us far afield, the point is that 
designers can, for any given customer grouping, determine \textit{which particular design trade-offs are ``worth it'' in terms of consumer group WTP}. Any set of 
candidate designs can be so assessed. Or, given a full-scale costing model, 
nonlinear optimization can determine the ``maximally profitable'' design, 
for a particular group, over the design space, just as in standard conjoint 
for traditional attributes. As an example of this last consideration, the 
model suggests that the four groups also have substantial heterogeneity in 
terms of WTP for additional gas mileage. As calculated earlier, the median 
value for the entire consumer set was {\$}2410, but the last row of Table 12 
suggests this is quite heterogeneous: {\$}970, {\$}1,730, {\$}6,290, 
{\$}3,110 for Groups 1-4, respectively. One might have guessed that the 
``cares about design'' Group 4 had high WTP overall, but they are only about 
average in WTP for additional gas mileage. Rather, Group 1, as suggested by 
the initial clustering, is an outlier in this regard, with each mile per 
gallon valued at only {\$}970, well under the actual savings over the 
typical lifetime of a car, and Group 4 at the other extreme, perhaps 
suggesting a distaste for combustion-based drivetrain technology. 

\section{Conclusion and Future Directions}
\label{subsec:mylabel9}
Preference elicitation is among the great success stories of experimental 
and statistical methodology resolving central problems in marketing. As 
evidenced by widespread adoption throughout the world over the last four 
decades (Sattler and Hartmann 2008), conjoint methods in particular can 
currently be deployed, using web-based tools, by practicing managers, with a 
low upfront burden in selecting optimal stimuli sets and backend estimation 
technologies. For example, Sawtooth's Discover allows product designers to 
specify attributes and levels, with subsequent ``heavy lifting'' -- 
orthogonal designs, choice-based stimuli, and Bayesian estimation -- handled 
seamlessly in the background. Yet even the best current implementations of 
conjoint founder on the shoals of visual design: while adjectival labels 
(e.g., sporty, bold, posh, etc.) and pre-generated 2D imagery can easily be 
included as categorical stimuli and covariates, and help directionally 
identify ``what consumers are looking for,'' they neither allow consumers to 
converge on specific designs that appeal to them nor designers to focus 
solely on \textit{the design space}, rather than pre-rendered descriptions or depictions of that 
space. 

This paper proposes what we believe to be the first comprehensive approach 
to the visual design problem, one that leverages both the sort of product 
topology modeling common in engineering and rapid, scalable machine learning 
algorithms to interweave with state-of-the-art preference elicitation 
methods developed in marketing. The resulting hybrid, using bi-level 
adaptive techniques and manipulable, real-time rendered imagery, can be 
deployed using standard web-based protocols to zero in on each consumer's 
preferred product design along with the sorts of attributes used 
traditionally in conjoint. The approach eschews descriptions or pre-set 
depictions of any sort, allowing post-hoc processing of individual-level 
data to determine trade-offs between common attributes like product price 
and visual design elements, as well as against design overall.

Our empirical analysis focuses deliberately on automotive design, as this is 
among the most complex durables that a consumer ever purchases: it is 
high-involvement, requires many trade-offs, and choices are deliberative. 
Cars are, notoriously, among the most design-intensive of all products. To 
take but a few ill-fated examples, Ford lavished over {\$}6Bn and several 
years on the design of its Mondeo ``world car'', Pontiac still stings over 
the Aztek, and the Edsel is the stuff of legend. Because the overwhelming 
majority of widely-deployed durable products involve computer-aided design 
(CAD), engineers can readily provide low-to-moderate dimensional product 
form representations for use in generating real-time 3D models for use 
within the method, with scant additional mediation by marketers or 
econometricians.

There are several ways to broaden the present approach, and perhaps to 
achieve even greater scalability. In the former category is including 
dedicated costing models and exogenous constraints on production 
feasibility, both of which are critical in real-world design. Michalek et 
al. (2011) achieved the latter for a small durable in a heterogeneous 
market, and is in fact not especially difficult to incorporate into a 
machine learning approach, using both soft (cost) and hard (feasibility) 
constraints. That is, a designer or engineer could designate the subspace of 
``buildable'' configurations and their attendant costs, so that either 
generated candidate 3D designs remained within its boundaries, or finalized 
products emerged from constrained or penalized optimization.

Our empirical example included a 19-dimensional design space that spanned a 
large swath of in-production consumer sedans, but hardly encompassed the 
full spectrum of passenger vehicles, let alone every aesthetic component 
thereof. While larger design spaces are possible, the main impediments are 
computational speed, efficient solution / generation of designs, and 
consumer willingness to participate in long conjoint tasks. One possibility 
takes a cue from Toubia et al. (2013), e.g., a look-up table could allow for 
rapid, adaptive query generation, as opposed to dedicated query-generation 
algorithms operating over a continuous (form or function) space. While 
consumer fatigue is a real issue for large number of ``optimized'' 
attributes, it can be mitigated through various experimental design and 
fusion-based techniques (Molin {\&} Timmermans 2009). 

For all their efficiency, adaptive algorithms could lead -- either as 
intermediate or final choices -- to ``undesirable'' designs, e.g., too 
costly for the manufacturer or seemingly odd to the consumer, especially in 
very large design spaces. Such algorithms typically involve some degree of 
hand-tuning, but a promising approach uses machine learning tools to weed 
out designs far away from those that have been either produced, or liked, 
conditional on an appropriate training corpus (Burnap et al. 2019). Such an 
approach would also help avoid ``wasting'' questions on stimuli unlikely to 
be selected, over and above prior responses in the design survey being 
administered.

These are all implementable improvements that could readily lead to crowdsourced, real-time,  manufacturer-feasible design optimization. The present method nevertheless demonstrates that judiciously-chosen query and estimation techniques, coupled with existing product topology models, render visual design feasible, without prior, analyst-supplied preconceptions of underlying design attributes. 

Appendices of the paper can be found \href{http://designinformaticslab.github.io/_papers/MKSC2019_Appendices.pdf}{here}.

\newpage 

\section{References}
{\setstretch{0.98}
\begin{hangparas}{.25in}{1}
Abernethy, Jacob, Theodoros Evgeniou, Olivier Toubia, J-P Vert. 2008. 
Eliciting consumer preferences using robust adaptive choice questionnaires. 
\textit{Knowledge and Data Engineering, IEEE Transactions on} \textbf{20} (2) 145--155.

Arora, Neeraj, Joel Huber. 2001. Improving parameter estimates and model 
prediction by aggregate customization in choice experiments. \textit{Journal of Consumer Research} \textbf{28} 
(2) 273--283.

Bloch, Peter H. 1995. Seeking the ideal form: product design and consumer 
response. \textit{The Journal of Marketing} 16--29.

Brazell, Jeff D., Christopher G. Diener, Ekaterina Karniouchina, William L. 
Moore, V\'{a}lerie S\'{e}verin, and Pierre-Francois Uldry. 2006. The 
no-choice option and dual response choice designs.~\textit{Marketing Letters}~\textbf{17} (4) 255-268. 

Burnap, Alex, John R. Hauser, and Artem Timoshenko. 2019. Design and 
Evaluation of Product Aesthetics: A Human-Machine Hybrid 
Approach.~\textit{Available at SSRN 3421771}.

Burnap, Alexander, Ye Liu, Yanxin Pan, Honglak Lee, Richard Gonzalez, and 
Panos Y. Papalambros. 2016. Estimating and exploring the product form design 
space using deep generative models. In~\textit{ASME 2016 International Design Engineering Technical Conferences and Computers and Information in Engineering Conference}, pp. V02AT03A013.

Chang, Chih-Chung, Chih-Jen Lin. 2011. LIBSVM: a library for support vector 
machines. \textit{ACM Transactions on Intelligent Systems and Technology (TIST)} \textbf{2} (3) 27.

Chapelle, O, SS Keerthi. 2010. Efficient algorithms for ranking with SVMs. 
\textit{Information Retrieval} \textbf{13} (3) 201--215.

Chapelle, Olivier, et al. 2004. A machine learning approach to conjoint 
analysis. \textit{Advances in neural information processing systems}. 257--264.

ClearVoice. 2014. ClearVoice research. http://www.clearvoiceresearch.com.

Cortes, Corinna, Vladimir Vapnik. 1995. Support-vector networks. \textit{Machine learning} 
\textbf{20} (3) 273--297.

Creusen, Marielle EH, Jan PL Schoormans. 2005. The different roles of 
product appearance in consumer choice*. \textit{Journal of product innovation management} \textbf{22} (1) 63--81.

Cristianini, Nello, John Shawe-Taylor. 2000. \textit{An introduction to support vector machines and other kernel-based learning methods}. Cambridge university press.

Cui, Dapeng, David Curry. 2005. Prediction in marketing using the support 
vector machine. \textit{Marketing Science} \textbf{24} (4) 595--615.

Dotson, Jeffrey P, Mark A Beltramo, Eleanor McDonnell Feit, Randall C Smith. 
2019. Modeling the Effect of Images on Product Choices. \textit{Available at SSRN 2282570} .

Dzyabura, Daria, John R Hauser. 2011. Active machine learning for 
consideration heuristics. \textit{Marketing Science}~\textbf{30} (5) 801-819.

Evgeniou, Theodoros, Constantinos Boussios, Giorgos Zacharia. 2005. 
Generalized robust conjoint estimation. \textit{Marketing Science} \textbf{24} (3) 415--429.

Evgeniou, Theodoros, Massimiliano Pontil, Olivier Toubia. 2007. A convex 
optimization approach to modeling consumer heterogeneity in conjoint 
estimation. \textit{Marketing Science} \textbf{26} (6) 805--818.

Fan, R.E., P.H. Chen, C.J. Lin. 2005. Working set selection using second 
order information for training support vector machines. \textit{The Journal of Machine Learning Research} \textbf{6} 
1889--1918.

Feit, Eleanor McDonnell, Pengyuan Wang, Eric T Bradlow, Peter S Fader. 2013. 
Fusing aggregate and disaggregate data with an application to multiplatform 
media consumption.~\textit{Journal of Marketing Research}~\textbf{50} (3) 348-364.

Feit, Eleanor McDonnell, Mark A Beltramo, Fred M Feinberg. 2010. Reality 
check: Combining choice experiments with market data to estimate the 
importance of product attributes. \textit{Management Science} \textbf{56} (5) 785--800.

Fixson, Sebastian K. 2007. Modularity and commonality research: past 
developments and future opportunities.~\textit{Concurrent Engineering}~\textbf{15} (2) 85-111.

Gramlich, Jacob, 2010. Gas prices, fuel efficiency, and endogenous product 
choice in the US automobile industry. Working paper, Georgetown 
University.

Green, Paul E, Venkat Srinivasan. 1990. Conjoint analysis in marketing: new 
developments with implications for research and practice.~\textit{The Journal of Marketing} \textbf{54 }(4) 
3-19.

Green, Paul E, Abba M Krieger, Yoram Wind. 2001. Thirty years of conjoint 
analysis: Reflections and prospects.~\textit{Interfaces},~\textbf{31} (3) S56-S73.

Greene, David L. 2010.~How consumers value fuel economy: A literature 
review. No. EPA-420-R-10-008.

Halme, Merja, Markku Kallio. 2011. Estimation methods for choice-based 
conjoint analysis of consumer preferences.~\textit{European Journal of Operational Research}, \textbf{214} (1) 160-167.

Helfand, Gloria, Ann Wolverton. 2009. Evaluating the consumer response to 
fuel economy: A review of the literature. National Center for Environmental 
Economics Working Paper 09-04.

Hsiao, S.W., Liu, M.C., 2002. A morphing method for shape generation and 
image prediction in product design.~\textit{Design studies},~\textbf{23} (6) 533-556.

Huang, Dongling, and Lan Luo 2016 Consumer preference elicitation of complex 
products using fuzzy support vector machine active learning."~\textit{Marketing Science}~\textbf{35} 
(30) 445-464.

Huber, Joel, and Klaus Zwerina. 1996. The importance of utility balance in 
efficient choice designs.~\textit{Journal of Marketing research}~\textbf{33} (3) 307-317. 

Joachims, Thorsten. 2002. Optimizing search engines using clickthrough data. 
\textit{Proceedings of the eighth ACM SIGKDD international conference on Knowledge discovery and data mining}. ACM, 133--142.

Kelly, Jarod C., Pierre Maheut, Jean-Fran\c{c}ois Petiot, Panos Y. 
Papalambros. 2011. Incorporating user shape preference in engineering design 
optimisation. \textit{Journal of Engineering Design} \textbf{22} (\ref{eq9}) 627--650.

Kim, H.J., Park, Y.H., Bradlow, E.T. and Ding, M. 2014. PIE: a holistic 
preference concept and measurement model.~\textit{Journal of Marketing Research},~\textbf{51 }(3) 335-351.

Kotler, Philip G, Alexander Rath. 1984. Design: A powerful but neglected 
strategic tool. Journal of Business Strategy \textbf{5} (2) 16--21.

Krishna, Aradhna, 2012. An integrative review of sensory marketing: Engaging 
the senses to affect perception, judgment and behavior.~\textit{Journal of Consumer Psychology},~\textbf{22 }(3) 
332-351.

Lai, Hsin-Hsi, Yu-Ming Chang, Hua-Cheng Chang. 2005. A robust design 
approach for enhancing the feeling quality of a product: a car profile case 
study. \textit{International Journal of Industrial Ergonomics} \textbf{35} (5) 445--460.

Landwehr, Jan R, Aparna A Labroo, Andreas Herrmann. 2011. Gut liking for the 
ordinary: Incorporating design fluency improves automobile sales forecasts. 
\textit{Marketing Science} \textbf{30} (3) 416--429.

Lenk, Peter J, Wayne S DeSarbo, Paul E Green, Martin R Young. 1996. 
Hierarchical bayes conjoint analysis: recovery of partworth heterogeneity 
from reduced experimental designs. \textit{Marketing Science} \textbf{15} (2) 173--191.

Lugo, Jos\'{e} E, Stephen M Batill, Laura Carlson. 2012. Modeling product 
form preference using gestalt principles, semantic space, and kansei. \textit{ASME 2012 International Design Engineering Technical Conferences and Computers and Information in Engineering Conference}. 
American Society of Mechanical Engineers, 529--539.

MacDonald, Erin, Alexis Lubensky, Bryon Sohns, Panos Y Papalambros. 2009. 
Product semantics and wine portfolio optimisation. \textit{International Journal of Product Development} \textbf{7} (1) 73--98.

Michalek, Jeremy J, Fred M Feinberg, Panos Y Papalambros. 2005. Linking 
marketing and engineering product design decisions via analytical target 
cascading."~\textit{Journal of Product Innovation Management}~\textbf{22} (1) 42-62.

Michalek, J. J., Ebbes, P., Adig\"{u}zel, F., Feinberg, F. M., {\&} 
Papalambros, P. Y. 2011. Enhancing marketing with engineering: Optimal 
product line design for heterogeneous markets.~\textit{International Journal of Research in Marketing},~\textbf{28} (1), 1-12.

Molin, Eric JE, and Harry JP Timmermans. 2009. Hierarchical information 
integration experiments and integrated choice experiments.~\textit{Transport reviews},~\textbf{29} (5) 
635-655.

Netzer, Oded, Olivier Toubia, Eric T Bradlow, Ely Dahan, Theodoros Evgeniou, 
Fred M Feinberg, Eleanor M Feit, Sam K Hui, Joseph Johnson, John C Liechty, 
et al. 2008. Beyond conjoint analysis: Advances in preference measurement. 
\textit{Marketing Letters} \textbf{19} (3-4) 337--354.

Gunay Orbay, Luoting Fu, and Levent Burak Kara. 2015. Deciphering the 
influence of product shape on consumer judgments through geometric 
abstraction. \textit{J. Mechanical Design} \textbf{137} (8) 081103.

Orsborn, Seth, Jonathan Cagan, Peter Boatwright. 2009. Quantifying aesthetic 
form preference in a utility function. \textit{Journal of Mechanical Design} \textbf{131} (6) 061001.

Orsborn, Seth, Jonathan Cagan. 2009. Multiagent shape grammar 
implementation: automatically generating form concepts according to a 
preference function. \textit{Journal of Mechanical Design} \textbf{131} (12) 121007.

Osugi, Thomas, Deng Kim, and Stephen Scott. ``Balancing exploration and 
exploitation: A new algorithm for active machine learning.'' In~\textit{Data Mining, Fifth IEEE International Conference on}, pp. 8-pp. 
IEEE, 2005.

Pan, Yanxin, Alexander Burnap, Jeffrey Hartley, Richard Gonzalez, and Panos 
Y. Papalambros. 2017. Deep design: Product aesthetics for heterogeneous 
markets. In~\textit{Proceedings of the 23rd ACM SIGKDD International Conference on Knowledge Discovery and Data Mining}, 1961-1970..

Qian, Yi, Hui Xie. 2013. Which brand purchasers are lost to counterfeiters? 
An application of new data fusion approaches.~\textit{Marketing Science}~\textbf{33} (3) 437-448.

Reid, Tahira N, Bart D Frischknecht, Panos Y Papalambros. 2012. Perceptual 
attributes in product design: Fuel economy and silhouette-based perceived 
environmental friendliness tradeoffs in automotive vehicle design. \textit{Journal of Mechanical Design} 
\textbf{134} (4) 041006.

Reid, Tahira N, Erin F MacDonald, Ping Du. 2013. Impact of product design 
representation on customer judgment. \textit{Journal of Mechanical Design} \textbf{135} (\ref{eq9}) 091008.

Ren, Yi, Panos Y Papalambros. 2011. A design preference elicitation query as 
an optimization process. \textit{Journal of Mechanical Design} \textbf{133} (11) 111004.

Rossi, Peter E, Greg M Allenby. 2003. Bayesian statistics and marketing. 
\textit{Marketing Science} \textbf{22} (3) 304--328.

Rossi, Peter E, Greg M Allenby, Robert E McCulloch. 2005. \textit{Bayesian statistics and marketing}. J. Wiley {\&} 
Sons.

Sattler, Henrik, Adriane Hartmann. 2008. Commercial use of conjoint 
analysis. In~\textit{Operations management in theorie und praxis}, pp. 103-119. Gabler.

Settles, Burr. 2010. Active learning literature survey. \textit{University of Wisconsin, Madison} \textbf{52} 55--66.

Sonnier, Garrett, Andrew Ainslie, Thomas Otter. 2007. Heterogeneity 
distributions of willingness-to-pay in choice models.~\textit{Quantitative Marketing and Economics}~\textbf{5} (3) 
313-331.

Sylcott, Brian, Jonathan Cagan, Golnaz Tabibnia. 2013a. Understanding 
consumer tradeoffs between form and function through metaconjoint and 
cognitive neuroscience analyses. \textit{Journal of Mechanical Design} \textbf{135} (10) 101002.

Sylcott, Brian, Jeremy J Michalek, Jonathan Cagan. 2013b. Towards 
understanding the role of interaction effects in visual conjoint analysis. 
\textit{ASME 2013 International Design Engineering Technical Conferences}. American Society of Mechanical Engineers, V03AT03A012.

Sylcott, Brian, Seth Orsborn, and Jonathan Cagan. 2016. The effect of 
product representation in visual conjoint analysis.~\textit{Journal of Mechanical Design}~\textbf{138} (10) 
101104.

Tong, S., D. Koller. 2002. Support vector machine active learning with 
applications to text classification. \textit{The Journal of Machine Learning Research} \textbf{2} 45--66.

Toubia, Olivier, Theodoros Evgeniou, John Hauser. 2007. \textit{Oxford Handbook of Innovation}, chap. 
Optimization-Based and Machine-Learning Methods for Conjoint Analysis: 
Estimation and Question Design. Springer, New York.

Toubia, Olivier, John Hauser, Rosanna Garcia. 2007. Probabilistic polyhedral 
methods for adaptive choice-based conjoint analysis: Theory and application. 
\textit{Marketing Science} \textbf{26} (5) 596--610.

Toubia, Olivier, Laurent Flor\`{e}s. 2007. Adaptive idea screening using 
consumers. \textit{Marketing Science} \textbf{26} (3) 342--360.

Toubia, Olivier, John R Hauser, Duncan I Simester. 2004. Polyhedral methods 
for adaptive choice-based conjoint analysis. \textit{Journal of Marketing Research} \textbf{41} (1) 116--131.

Toubia, Olivier, Eric Johnson, Theodoros Evgeniou, and Philippe Delqui\'{e}. 
2013. Dynamic experiments for estimating preferences: An adaptive method of 
eliciting time and risk parameters.~\textit{Management Science}~\textbf{59} (3) 613-640.

Toubia, Olivier, Duncan I Simester, John R Hauser, Ely Dahan. 2003. Fast 
polyhedral adaptive conjoint estimation. \textit{Marketing Science} \textbf{22} (3) 273--303.

Tseng, Ian, Jonathan Cagan, Kenneth Kotovsky. 2012. Concurrent optimization 
of computationally learned stylistic form and functional goals. \textit{Journal of Mechanical Design} 
\textbf{134} (11) 111006.

\end{hangparas}

} 

\end{document}